\documentclass[twocolumn,preprintnumbers,amsmath,amssymb]{revtex4}
\usepackage{amsmath}
\usepackage[dvips]{graphicx}
\usepackage{graphics}
\usepackage{amssymb}

\newcommand{\ppt }{P_{\alpha}^{(+)}(t)}
\newcommand{\dea }{\Delta\epsilon^{\alpha}}

\newcommand{\del }{\Delta\epsilon^{L}}
\newcommand{\der }{\Delta\epsilon^{R}}
\newcommand{\dvl }{\Delta V^L}
\newcommand{\dvr }{\Delta V^R}

\newcommand{\br}{\mathbf{r}}

\newcommand{\iqa}{Q_{\alpha}}
\newcommand{\iqawbl}{Q^{\mathit{WBL}}_{\alpha}}
\newcommand{\irhod}{\rho_D(\br, t)}
\newcommand{\ila}{\Lambda^{\alpha}}
\newcommand{\irhoa}{\rho_\alpha}
\newcommand{\saa}{\Sigma^a_{\alpha}}
\newcommand{\sla}{\Sigma^<_{\alpha}}
\newcommand{\sga}{\Sigma^>_{\alpha}}
\newcommand{\ict}{\mathit{CT}}
\newcommand{\beq}{\begin{equation}}
\newcommand{\enq}{\end{equation}}
\newcommand{\be}{\begin{eqnarray}}
\newcommand{\en}{\end{eqnarray}}
\newcommand{\ka}{k_\alpha}
\newcommand{\pb}{p_\beta}
\newcommand{\eka}{\epsilon_{\ka}}
\newcommand{\grk}{g^r_{\ka}(t, \tau)}
\newcommand{\gak}{g^a_{\ka}(t, \tau)}
\newcommand{\glk}{g^<_{\ka}(t, \tau)}

\newcommand{\tslga}{\tilde{\Sigma}^{<,>}_{\alpha}}

\begin{document}

\title{Time-dependent density-functional theory for open systems}
\author{Xiao Zheng}
\author{Fan Wang}
\author{Chi Yung Yam}
\author{Yan Mo}
\author{GuanHua Chen}
\email{ghc@everest.hku.hk} \affiliation{Department of Chemistry,
The University of Hong Kong, Hong Kong, China}

\date{\today}


\begin{abstract}
By introducing the
self-energy density functionals for the dissipative interactions
between the reduced system and its environment, we develop a
time-dependent density-functional theory formalism based on an
equation of motion for the Kohn-Sham reduced single-electron
density matrix of the reduced system.
Two approximate schemes are proposed for the self-energy density
functionals, the complete second order approximation and the
wide-band limit approximation. A numerical method based on the
wide-band limit approximation is subsequently developed and
implemented to simulate the steady and transient
current through various realistic molecular devices. Simulation
results are presented and discussed.

\end{abstract}

\maketitle

\section {Introduction \label{introduction}}

Density-functional theory (DFT) has been widely used as a research
tool in condensed matter physics, chemistry, materials science,
and nanoscience. The Hohenberg-Kohn theorem~\cite{hk} lays the
foundation of DFT. The Kohn-Sham (KS) formalism~\cite{ks} provides
a practical solution to calculate the ground state properties of
electronic systems. Runge and Gross extended DFT further to
calculate the time-dependent properties and hence the excited
state properties of any electronic systems~\cite{tddft}. The
accuracy of DFT or time-dependent DFT (TDDFT) is determined by the
exchange-correlation (XC) functional. If the exact XC functional
were known, the KS formalism would have provided the exact ground
state properties, and the Runge-Gross extension, TDDFT, would have
yielded the exact time-dependent and excited states properties.
Despite their wide range of applications, DFT and TDDFT have been
mostly limited to isolated systems.

Many systems of current research interest are open systems. A
molecular electronic device is one such system. Simulations based
on DFT have been carried out on such devices~\cite{prllang,
prlheurich, jcpluo, langprb, prbguo, prbywt, jacsywt, jacsgoddard,
transiesta, jcpratner}. These simulations focus on steady-state
currents under bias voltages. Two types of approaches have been
adopted. One is the Lippmann-Schwinger formalism by Lang and
coworkers~\cite{langprb}. The other is the first-principles
nonequilibrium Green's function (NEGF) technique~\cite{prbguo,
prbywt, jacsywt, jacsgoddard, transiesta, jcpratner}. In both
approaches the KS Fock operator is taken as the effective
single-electron model Hamiltonian, and the transmission
coefficients are calculated within the noninteracting electron
model. The investigated systems are not in their ground states,
and applying ground state DFT formalism for such systems is only
an approximation~\cite{cpdatta}. DFT formalisms adapted for
current-carrying systems have also been proposed recently, such as
Kosov's KS equations with direct current~\cite{jcpkosov}, Kurth
\emph{et al.}'s~\cite{kurth1} and Zheng~\emph{et al.}'s \cite{zhengarx05}
TDDFT formulation, Cui \emph{et al.}'s
complete second-order quantum dissipation theory (CS-QDT)
formalism~\cite{csqdt-scba} and Burke \emph{et al.}'s KS master
equation including dissipation to phonons~\cite{prlburke}. In this
paper, we present a new DFT formalism for open electronic systems,
and use it to simulate the steady and transient currents through
molecular electronic devices. The first-principles formalism
depends only on the electron density function of the reduced
system.

As early as in 1981, Riess and M\"{u}nch~\cite{riess} discovered the
holographic electron density theorem which states that any nonzero
volume piece of the ground state electron density determines the
electron density of a molecular system. This is based on that the
electron density functions of atomic and molecular eigenfunctions
are real analytic away from nuclei. In 1999 Mezey extended the
holographic electron density theorem~\cite{mezey}. And in 2004
Fournais~\emph{et al.} proved again the real analyticity of the
electron density functions of any atomic or molecular
eigenstates~\cite{analyticity}. Therefore, for a time-independent
real physical system made of atoms and molecules, its electron
density function is real analytic (except at nuclei) when the system
is in its ground state, any of its excited eigenstates, or any state
which is a linear combination of finite number of its eigenstates;
and the ground state electron density on any finite subsystem
determines completely the electronic properties of the entire
system.

As for time-dependent systems, the issue was less clear until
recently we~\cite{openprb} were able to
establish a one-to-one correspondence between the electron density
function of any finite subsystem and the external potential field
which is real analytic in both $t$-space and $\mathbf{r}$-space.
For time-dependent real physical systems,
we have proved the following theorem:~\cite{openprb}

{\it Theorem:} If the electron density function of a real finite
physical system at $t_0$, $\rho(\mathbf{r},t_0)$, is real analytic
in $\mathbf{r}$-space, the corresponding wave function is
$\Phi(t_0)$, and the system is subjected to a real analytic (in
both $t$-space and $\mathbf{r}$-space) external potential field
$v(\mathbf{r},t)$, the time-dependent electron density function on
any finite subspace $D$, $\rho_D(\mathbf{r},t)$, has a one-to-one
correspondence with $v(\mathbf{r},t)$ and determines uniquely all
electronic properties of the entire time-dependent system.

According to the \emph{Theorem}, the electron density
function of any subsystem determines all the electronic properties
of the entire time-dependent physical system. This proves in
principle the existence of a TDDFT formalism for open electronic
systems. All one needs to know is the electron density function of
the reduced system.

This paper is organized as follows.
In Sec.~\ref{formalism} we describe a TDDFT formalism for open
electronic systems based on an equation of motion (EOM) for the
reduced single-electron density matrix. By utilizing the
holographic electron density theorem, the self-energy functionals
with explicit functional dependence on the electron density of the
reduced system are introduced, and thus a rigorous and efficient
first-principles formalism for the transient dynamics of any open
electronic system is established. Two approximate schemes, the
complete second order (CSO) approximation for the dissipative interaction
and the wide-band limit (WBL) approximation for the electrodes, are
proposed for the self-energy functionals in Sec.~\ref{formalism}.
To demonstrate the applicability of our first-principles formalism,
TDDFT calculations are carried out to simulate the transient and
steady current through realistic molecular devices. The detailed
numerical procedures and results are described in
Sec.~\ref{tddft-wbl}. Discussion and summary are given in
Sec.~\ref{summary}.

\section{First-principles formalism \label{formalism}}

\subsection{Equation of motion \label{eom-system}}

\begin{figure}
\includegraphics[scale=0.45]{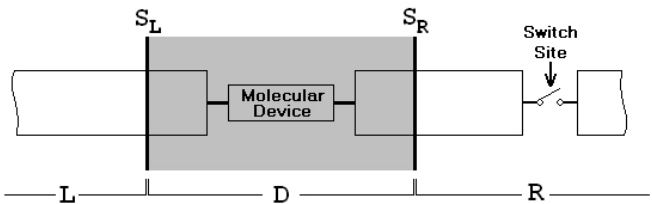}
\caption{\label{scheme} Schematic representation of the experimental
setup for quantum transport through a molecular device.}
\end{figure}

Fig.~\ref{scheme} depicts an open electronic system. Region $D$ is
the reduced system of our interests, and the electrodes $L$ and
$R$ are the environment. Altogether $D$, $L$ and $R$ form the
entire system. Taking Fig.~\ref{scheme} as an example, we develop
a practical DFT formalism for the open systems. Within the TDDFT
formalism, a closed EOM has been derived for the reduced
single-electron density matrix $\sigma(t)$ of the entire
system~\cite{ldmtddft}:
\begin{equation}\label{eom4sigma0}
    i\dot{\sigma}(t) = [h(t),\sigma(t)],
\end{equation}
where $h(t)$ is the KS Fock matrix of the entire system, and the
square bracket on the right-hand side (RHS) denotes a commutator.
The matrix element of $\sigma$ is defined as $\sigma_{ij}(t) =
\langle a^{\dagger}_{j}(t)\,a_{i}(t)\rangle$, where $a_{i}(t)$ and
$a^{\dagger}_{j}(t)$ are the annihilation and creation operators
for atomic orbitals $i$ and $j$ at time $t$, respectively. Fourier
transformed into frequency domain while considering linear
response only, Eq.~(\ref{eom4sigma0}) leads to the conventional
Casida's equation~\cite{casida}. Expanded in the atomic orbital
basis set, the matrix representation of $\sigma$ can be
partitioned as
\begin{equation}\label{matrixsigma}
    \sigma = \left[\begin{array}{lll}
    \sigma_{L}  & \sigma_{LD} & \sigma_{LR} \\
    \sigma_{DL} & \sigma_{D}  & \sigma_{DR} \\
    \sigma_{RL} & \sigma_{RD} & \sigma_{R}
    \end{array}\right],
\end{equation}
where $\sigma_{L}$, $\sigma_{R}$ and $\sigma_{D}$ represent the
diagonal blocks corresponding to the left lead $L$, the right lead
$R$ and the device region $D$, respectively; $\sigma_{LD}$ is the
off-diagonal block between $L$ and $D$; and $\sigma_{RD}$,
$\sigma_{LR}$, $\sigma_{DL}$, $\sigma_{DR}$ and $\sigma_{RL}$ are
similarly defined. The KS Fock matrix $h$ can be partitioned in
the same way with $\sigma$ replaced by $h$ in
Eq.~(\ref{matrixsigma}). Thus, the EOM for $\sigma_{D}$ can be
written as
\begin{eqnarray}\label{eom4sigmad0}
    i\dot{\sigma}_{D} &=& [h_{D},\sigma_{D}] + \sum_{\alpha=L,R}
    \left(h_{D\alpha}\sigma_{\alpha D}-\sigma_{D\alpha}
    h_{\alpha D}\right) \nonumber \\
    &=& [h_{D},\sigma_{D}] - i\sum_{\alpha=L,R}Q_{\alpha},
\end{eqnarray}
where $Q_{L}$ ($Q_{R}$) is the dissipation term due to $L$ ($R$).
With the reduced system $D$ and the leads $L/R$ spanned
respectively by atomic orbitals $\{l\}$ and single-electron states
$\{k_{\alpha}\}$, Eq.~(\ref{eom4sigmad0}) is equivalent to:
\begin{eqnarray}\label{eom4sigmad1}
    i\dot{\sigma}_{nm} &=& \sum_{l\in
    D}\,(h_{nl}\sigma_{lm}-\sigma_{nl}h_{lm}) - i\sum_{\alpha=L,R}
    Q_{\alpha,nm}, \label{eom4sigmad2} \\
    Q_{\alpha,nm} &=& i\sum_{k_{\alpha}\in\alpha}\big(h_{nk_{\alpha}}
    \sigma_{k_{\alpha}m}-\sigma_{nk_{\alpha}}
    h_{k_{\alpha}m}\big),\label{qterm0}
\end{eqnarray}
where $m$ and $n$ correspond to the atomic orbitals in region $D$;
$k_{\alpha}$ corresponds to an electronic state in the electrode
$\alpha$ ($\alpha = L$ or $R$). $h_{nk_{\alpha}}$ is the coupling
matrix element between the atomic orbital $n$ and the electronic
state $k_{\alpha}$. The transient current through the interfaces
$S_L$ or $S_R$ (see Fig.~\ref{scheme}) can be evaluated as
follows,
\begin{eqnarray}\label{jcurrent}
    J_{\alpha}(t) &=& -\int_{\alpha}d\mathbf{r}\,\frac{\partial}
    {\partial t}\rho(\mathbf{r},t) \nonumber \\
    &=& -\sum_{k_{\alpha}\in\alpha}\frac{d}
    {dt}\,\sigma_{k_{\alpha}k_{\alpha}}\!(t) \nonumber \\
    &=& i\sum_{l\in D}\sum_{k_{\alpha}\in\alpha}\big(
    h_{k_{\alpha}l}\,\sigma_{lk_{\alpha}} -
    \sigma_{k_{\alpha}l}\,h_{lk_{\alpha}}\big) \nonumber \\
    &=& -\sum_{l\in D}Q_{\alpha,ll}
    = -\mbox{tr}\big[Q_{\alpha}(t)\big].
\end{eqnarray}
Since the dissipation term $\iqa(t)$ is not known \emph{a priori},
Eq.~(\ref{eom4sigmad0}) is not self-closed. Therefore, at this
stage EOM~(\ref{eom4sigmad0}) cannot be solved straightforwardly
to obtain the transient dynamics of the reduced system $D$.

According to the holographic electron density theorem of
time-dependent physical systems, all physical quantities are
explicit or implicit functionals of the electron density in the
reduced system $D$, $\rho_{D}(\mathbf{r},t)$. $Q_{\alpha}$ of
Eq.~(\ref{eom4sigmad0}) is thus also a functional of
$\rho_{D}(\mathbf{r},t)$. Therefore, Eq.~(\ref{eom4sigmad0}) can
be recast into a formally closed form,
\begin{equation}\label{e4rdm}
  i \dot{\sigma}_D = \Big[h_{D}[t; \rho_{D}
  (\br, t)], \sigma_{D}\Big] - i\!\!\sum_{\alpha=L,R}
  \!\!Q_{\alpha}[t ;\rho_{D}(\br, t)].
\end{equation}
It would thus be much more efficient integrating Eq.~(\ref{e4rdm})
than solving Eq.~(\ref{eom4sigmad1}), provided that $\iqa[t;
\irhod]$ or its approximation is known. We therefore have a
practical formalism for any open electronic systems. Neglecting
$\iqa[t; \irhod]$ from Eq.~(\ref{e4rdm}) leads to the conventional
TDDFT formulation~\cite{ldmtddft} for the isolated reduced system,
while $Q_{\alpha}[t; \rho_{D}(\br, t)]$ accounts for the
dissipative interactions between $D$ and $L$ or $R$.
Eq.~(\ref{e4rdm}) is the TDDFT EOM for open electronic systems,
and is formally analogous to the master equations derived for the
system reduced density matrix in conventional QDT~\cite{qdt}.

Our formalism is similar in its form to one of our early works, in
which a dynamic mean-field theory for dissipative interacting
many-electron systems was developed~\cite{yokojima1,yokojima2}. An
EOM for the reduced single-electron density matrix was derived to
simulate the excitation and nonradiative relaxation of a molecule
embedded in a thermal bath. This is in analogy to our case
although our environment is actually a fermion bath instead of a
boson bath. More importantly, the number of electrons in the
reduced system is conserved in Refs.~\cite{yokojima1,yokojima2}
while in our case it is not.

Burke \emph{et al.} extended TDDFT to include electronic systems
interacting with phonon baths~\cite{prlburke}, they proved the
existence of a one-to-one correspondence between $v(\mathbf{r},t)$
and $\rho(\mathbf{r},t)$ under the condition that the dissipative
interactions (denoted by a superoperator $\mathcal{C}$ in
Ref.~\cite{prlburke}) between electrons and phonons are fixed. In
our case since the electrons can move in and out the reduced
system, the number of the electrons in the reduced system is not
conserved. In addition, the dissipative interactions can be
determined in principle by the electron density of the reduced
system. We do not need to stipulate that the dissipative
interactions with the environment are fixed as Burke \emph{et
al.}. And the only information we need is the electron density of
the reduced system. In the frozen DFT approach~\cite{warshel} an
additional kinetic energy functional term caused by the
environment was introduced to account for the
interaction between the system and the environment. This
additional term is included in $Q_{\alpha}[t; \rho_D(\br, t)]$ of
Eq.~(\ref{e4rdm}).

\subsection{The dissipation term $\iqa$ \label{q-term}}

The challenge now is to express $\iqa[t; \irhod]$. Based on the
Keldysh formalism~\cite{keldysh} and the analytical continuation
rules of Langreth~\cite{langreth}, $\iqa(t)$ can be calculated by
the NEGF formulation as described in Reference~\cite{prb94win}
(see Appendix~\ref{derivee4qt}, \emph{cf.} Eq.~(\ref{qterm0})):
\begin{eqnarray}
    Q_{\alpha,nm}(t)&=&-\sum_{l\in D}\int_{-\infty}^{\infty}d\tau
    \Big[\,G^{<}_{nl}(t,\tau)\Sigma^{a}_{\alpha,lm}(\tau,t)
    +\nonumber\\
    &&\,G^{r}_{nl}(t,\tau)\Sigma^{<}_{\alpha,lm}(\tau,t)
    + \mbox{H.c.} \Big], \label{e4qt}
\end{eqnarray}
where $G^r_D$ and $G^<_D$ are the retarded and lesser Green's
function of the reduced system $D$, and $\Sigma^a_{\alpha}$ and
$\Sigma^<_\alpha$ are the advanced and lesser self-energies due to
the lead $\alpha$ ($L$ or $R$), respectively. Combining
Eqs.~(\ref{jcurrent}) and (\ref{e4qt}), we obtain
\begin{eqnarray}\label{jqttddft}
    J_{\alpha}(t) &=& 2\Re\,\bigg\{\int_{-\infty}^{\infty}\!d\tau\,
    \,\mbox{tr}\Big[G^{<}_{D}(t,\tau)\Sigma^{a}_{\alpha}(\tau,t) +
    \nonumber\\
    && G^{r}_{D}(t,\tau)\Sigma^{<}_{\alpha}(\tau,t)\Big]\bigg\}.
\end{eqnarray}
Eq.~(\ref{jqttddft}) has been derived by Stefanucci and Almbladh~\cite{qttddft}
within the framework of TDDFT under the assumaption that the
partitioned~\cite{caroli} and partition-free~\cite{cini}
schemes are equivalent.

It is important to emphasize that Eq.~(\ref{e4qt}) is derived from
the initial ground state at $t = -\infty$ when the device region
and the leads are completely isolated, denoted by $\Phi_0$. This
corresponds to the partitioned scheme developed by Caroli~\emph{et
al.}~\cite{caroli}. The dissipation term $\iqa$ can also be
derived from the initial ground state at $t = t_0$ when the device
region and the leads are fully connected, denoted by $\Psi_0$, as
follows (see Appendix~\ref{app-iqa} for detailed derivations), \be
\label{e4qt2}
  Q_{\alpha, nm}(t) &=& \bigg\{ Q^0_{\alpha, nm}(t) -\! \sum_{l\in
  D} \! \int_{t_0^+}^t d\tau \Big[ G^{<}_{nl}(t,\tau)\Sigma^{a}_{\alpha,lm}
  (\tau,t) \nonumber\\
  && +\, G^{r}_{nl}(t,\tau)\Sigma^{<}_{\alpha,lm}(\tau,t) \Big] \bigg\}
  +\, \mbox{H.c.},
\en where $t_0^+$ is the time immediately after $t_0$, and the
first term on the RHS, $Q^0_{\alpha, nm}(t)$, arises due to the
initial couplings between the reduced system and the environment.
Eq.~(\ref{e4qt2}) thus follows the partition-free scheme proposed
by Cini~\cite{cini}, and its associated Green's functions and
self-energies are defined differently from those in
Eq.~(\ref{e4qt}).

Based on Gell-Mann and Low theorem~\cite{gml}, in most cases
$\Psi_0$ can be reached from $\Phi_0$ by adiabatically turning on
the couplings between the device and the leads from $t = -\infty$
to $t_0$. In these circumstances, the partitioned and
partition-free schemes are formally equivalent, since the history
of the couplings between the device and leads only determines
$\Psi_0$ and its corresponding electron density function
$\rho(\br, t_0)$, and does not influence the dynamic response of
the reduced system to external potentials after $t_0$ explicitly.
In few cases where the turn-on of the couplings results in an
excited eigenstate at $t_0$, Eq.~(\ref{e4qt}) is only an
approximation for the $\iqa$ derived from $\Psi_0$ in the
partition-free scheme, and in principle we need to resort to
Eq.~(\ref{e4qt2}).

\subsection{Solution for steady-state current \label{sscurrent}}

In cases where steady states can be reached, the
system-bath coupling, $\Gamma^{k_{\alpha}}_{nm}(t,\tau) \equiv
h_{nk_ {\alpha}}(t)\,h_{k_{\alpha}m}(\tau)$, becomes
asymptotically time-independent as $t,\tau\rightarrow +\infty$.
The Green's functions and self-energies for the reduced system $D$
rely simply on the difference of the two
time-variables~\cite{qttddft}, \emph{i.e.}, $G_D(t, \tau) \sim
G_D(t - \tau)$ and $\Sigma(t, \tau) \sim \Sigma(t - \tau)$, and
thus we have
\be \label{expandedglra}
   G^{<}_{nm}(t,\tau) &=& \sum_{p,q\in D}\int_{-\infty}
   ^{\infty}dt_{1}\int_{-\infty}^{\infty}dt_{2}\,G^{r}_{np}
   (t,t_{1})\nonumber \\
   &&\times\,\Sigma^{<}_{pq}(t_{1},t_{2})\,G^{a}_{qm}(t_{2},\tau)\nonumber\\
   &=& i\sum_{p,q\in D}\sum_{\alpha=L,R}\sum_{l_{\alpha}\in\alpha}f^
   {\alpha}_{l}\nonumber \\
   &&\times\left[\int_{-\infty}^{\infty}dt_{1}\textnormal{e}^
   {-i\epsilon_{l}^{\alpha}t_{1}}G^{r}_{np}(t-t_{1})\right]\Gamma
   ^{l_{\alpha}}_{pq}\nonumber\\
   &&\times\left[\int_{-\infty}^{\infty}dt_{2}\textnormal{e}
   ^{i\epsilon_{l}^{\alpha}t_{2}}G^{a}_{qm}(t_{2}-\tau)\right]
   \nonumber\\
   &=& i\sum_{p,q\in D}\sum_{\alpha=L,R}\sum_{l_{\alpha}\in\alpha}
   f^{\alpha}_{l}\textnormal{e}^{-i\epsilon^{\alpha}_{l}(t-\tau)}
   \nonumber \\
   && \times\,G^{r}_{np}(\epsilon^{\alpha}_{l})\,\Gamma^{l_{\alpha}}_{pq}\,
   G^{a}_{qm}(\epsilon^{\alpha}_{l}),\label{g01}
\en %
\be
   G^{r,a}_D(\epsilon) &=& \left[\epsilon I - h_D(\infty) - \Sigma^{r,a}(\epsilon)
   \right]^{-1}, \label{g02}\\
   \Sigma^{r,a}_{nm}(\epsilon) &=&
   \sum_{\alpha=L,R}\sum_{l\in\alpha}\,
   \Gamma^{l_{\alpha}}_{nm}\left(\epsilon-\epsilon_{l}^{\alpha}\pm
   i\delta\right)^{-1}, \label{g03}
\en
where $I$ is an identity matrix, $\delta$ is an infinitesimal
positive number, and $f^\alpha_l$ is the occupation number of the
single-electron state $l_\alpha$ of the isolated lead $\alpha$
($L$ or $R$). The steady-state current can thus be explicitly
expressed by combining Eqs.~(\ref{g01})$-$(\ref{g03}),
\begin{eqnarray}\label{jss}
    J_{L}(\infty) &=& -J_{R}(\infty) \nonumber \\
    &=& -\sum_{n\in D}Q_{L,nn}(\infty) \nonumber\\
    &=& 2\pi\,\Bigg\{\sum_{k\in L}f^{L}_{k}\sum_{l\in R}\delta
    (\epsilon^{R}_{l}-\epsilon^{L}_{k})\nonumber \\
    && \times\,\mbox{tr}\Big[G^{r}_
    {D}(\epsilon^{L}_{k})\,\Gamma^{l_{R}}\,G^{a}_{D}(\epsilon^{L}
    _{k})\,\Gamma^{k_{L}}\Big] \nonumber \\
    &&-\sum_{l\in R}f^{R}_{l}\sum_{k\in L}
    \delta(\epsilon^{L}_{k}-\epsilon^{R}_{l})\nonumber \\
    &&\times\,\mbox{tr}\Big[
    G^{r}_{D}(\epsilon^{R}_{l})\,\Gamma^{l_{R}}\,G^{a}_{D}
    (\epsilon^{R}_{l})\,\Gamma^{k_{L}}\Big]\Bigg\} \nonumber\\
    &=&\int\left[f^{L}(\epsilon)-f^{R}(\epsilon)\right]T(\epsilon)
    \,d\epsilon,\\
    T(\epsilon)&=&2\pi\,\eta_{L}\eta_{R}\,\mbox{tr}
    \Big[G^{r}_{D}(\epsilon)\Gamma^{R}(\epsilon)G^{a}_{D}
    (\epsilon)\Gamma^{L}(\epsilon)\Big].\label{tofe}
\end{eqnarray}
Here $T(\epsilon)$ is the KS transmission coefficient,
$f^{\alpha}(\epsilon)$ is the Fermi distribution function, and
$\eta_{\alpha}(\epsilon) \equiv \sum_{k\in\alpha}\delta(\epsilon
-\epsilon ^{\alpha}_{k})$ is the density of states (DOS) for the
lead $\alpha$ ($L$ or $R$). Eq.~(\ref{jss}) appears formally
analogous to the Landauer formula~\cite{bookdatta, landauer}
adopted in the conventional DFT-NEGF formalism~\cite{prbguo,
jacsywt}. However, to obtain the correct steady current, the
nonequilibrium effects need to be properly accounted for. This may
be accomplished by substituting the asymptotic values of the TDDFT
XC potential for the ground state DFT counterpart in
Eq.~(\ref{jss}).

\subsection{Self-energy functionals \label{selffunc}}

Due to its convenience for practical implementation,
Eq.~(\ref{e4qt}) is adopted in our formalism. The Green's
functions $G_D^r$ and $G_D^<$ in Eq.~(\ref{e4qt}) can be
calculated via the following EOMs if $\saa$ and $\sla$ are known,
\begin{eqnarray}
   i\frac{\partial\,G^{r}_{nm}(t, \tau)}{\partial t} &=&
   \delta(t - \tau)\,\delta_{nm} + \sum_{l\in D}
   h_{nl}(t)\,G^{r}_{lm}(t, \tau) \nonumber \\
   &&  + \sum_{l\in D}\int_{-\infty}^{\infty}
   d\bar{t}\,\Sigma^{r}_{nl}(t, \bar{t})\,G^{r}_{lm}(\bar{t},
   \tau),   \label{eom4gretard} \\
   i\frac{\partial\,G^{<}_{nm}(t, \tau)}{\partial t}
   &=& \sum_{l\in D}\int_{-\infty}^{\infty}\!\! d\bar{t}\,
   \Big[\Sigma^{<}_{nl}(t, \bar{t})\,G^{a}_{lm}(\bar{t}, \tau)
   + \Sigma^{r}_{nl}(t, \bar{t}) \nonumber\\
   && \times\,G^{<}_{lm}(\bar{t}, \tau)
   \Big] + \sum_{l\in D}h_{nl}(t)\,G^{<}_{lm}(t, \tau),
   \label{eom4gless}
\end{eqnarray}
where $\Sigma^r = \sum_{\alpha=L,R}(\saa)^\dag$, $\Sigma^< =
\sum_{\alpha=L,R}\sla$, and $G^a_D = (G^r_D)^\dag$. The key
quantities for the evaluation of $\iqa[t; \irhod]$ are thus the
self-energies $\saa$ and $\sla$. According to our \emph{Theorem},
$\saa$ and $\sla$ are in principle functionals of $\irhod$.
Therefore, instead of finding $\iqa[t; \irhod]$ directly, we need
now seek for the density functionals $\saa[\tau, t; \irhod]$ and
$\sla[\tau, t; \irhod]$. By their definitions, the self-energy
terms have explicit functional dependence on the electron density
function of the entire system, $\rho = (\rho_D, \irhoa)$:
\begin{eqnarray}
   \saa[\tau, t; \rho] &\equiv& i\,\vartheta(t-\tau)\,h[\tau; \rho]
   \exp \left\{ i\!\int_\tau^t h_{\alpha}[\bar{t};
   \rho_{\alpha}] d\bar{t}\right\} \nonumber \\
   && \times \,h[t;\rho], \label{selfa} \\
   \sla[\tau, t; \rho] &\equiv& i \,h[\tau; \rho]\,
   f^{\alpha}\!\left(h_\alpha[-\infty; \rho_\alpha]\right)\nonumber \\
   && \times \exp\left\{ i\!\int_\tau^t h_{\alpha}[\bar{t};
   \rho_{\alpha}] d\bar{t}\right\}\,h[t;\rho], \label{selfl}
\end{eqnarray}
where $\irhoa$ is the electron density function in the lead
$\alpha$, $h_\alpha$ is the KS Fock matrix of the isolated lead
$\alpha$, and $f^\alpha$ is the Fermi distribution function for
$\alpha$ ($L$ or $R$). Based on our~\emph{Theorem}, $\irhoa$ are
determined uniquely by $\rho_D$ via a certain
continuation~($\ict$) operation, \emph{i.e.},
\begin{eqnarray}
   \irhod & \stackrel{\ict}{\longrightarrow} &
   \rho_\alpha(\br, t), \label{ct1} \\
   \rho_\alpha(\br, t) &=&
   \rho^{\ict}_\alpha[\br, t; \irhod]. \label{ct2}
\end{eqnarray}
We obtain thus the following functionals,
\begin{eqnarray}
   \saa(\tau, t) & = & \saa\left[\tau, t; \rho_D, \irhoa^\ict[\rho_D]\right],
   \label{saa-exact} \\
   \sla(\tau, t) & = & \sla\left[\tau, t; \rho_D, \irhoa^\ict[\rho_D]\right].
   \label{sla-exact}
\end{eqnarray}
Note that the $\ict$ operation is case dependent, and often
approximate in practice. For the system depicted in
Fig.~\ref{scheme}, the $\ict$ operation from $\rho_D$ to $\irhoa$
may be approximated by a translation over repeating unit cells if
the bulk electrodes are periodic, \emph{i.e.},
\begin{equation}
   \irhoa(\br, t) = \irhoa^\ict[\rho_D] \approx \rho_D(\br + N \mathbf{R},
   t), \label{ct-approx}
\end{equation}
where $t=0$ refers to the initial time when the entire connected
system is in its ground state, $\mathbf{R}$ is the base vector
perpendicular to the interface $S_\alpha$ for the lead $\alpha$,
and $N$ denotes an integer which makes the translated vector $\br
+ N \mathbf{R}$ to be inside the reduced system $D$ as well as
near the interfaces $S_\alpha$. To ensure the accuracy of such an
approximate $\ict$ operation, it is vital to include enough
portions of electrodes into the region $D$, so that the electron
density function near the interfaces $S_\alpha$ takes correctly
the bulk values.

Of course, there could be cases that the approximate
$\irhoa^\ict[\rho_D]$ may deviate drastically from their exact
values some distance away from the boundary. Usually $\saa$ and
$\sla$ depend mostly on the electron density near the boundary
where the approximate $\irhoa^\ict[\rho_D]$ agree best with the
correct $\irhoa$. The resulting $\saa[\rho_D,
\irhoa^\ict[\rho_D]]$ and $\sla[\rho_D, \irhoa^\ict[\rho_D]]$ thus
provide reasonable approximations for their exact counterparts.
For cases where the self-energies happen to rely heavily on
$\irhoa$ far away from $D$, the approximated $\ict$ breaks down,
and our method fails to be applicable.

\begin{figure}
\includegraphics[scale=0.4]{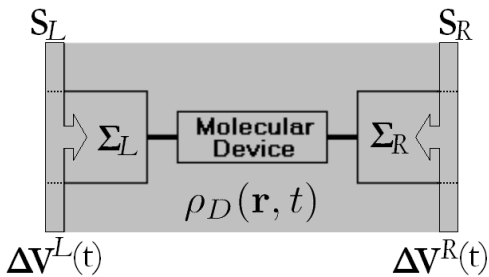}
\caption{\label{device} The molecular device region $D$ is subject
to the boundary conditions $\dvl(t)$ and $\dvr(t)$ at the
interfaces $S_L$ and $S_R$. The interactions between the region
$D$ and the lead $L$ and $R$ are accounted for by the self-energy
functionals $\Sigma_L$ and $\Sigma_R$, respectively. }
\end{figure}

Given $\saa[\rho_D]$ and $\sla[\rho_D]$ how do we solve the
EOM~(\ref{e4rdm}) in practice? Again take the molecular device
shown in Fig.~\ref{scheme} as an example. We focus on the reduced
system $D$ as depicted in Fig.~\ref{device}, and integrate the
EOM~(\ref{e4rdm}) directly by satisfying the boundary conditions
at $S_{L}$ and $S_{R}$. $\Delta V^L(t)$ and $\Delta V^R(t)$ are
the bias voltages applied on $L$ and $R$, respectively, and serve
as the boundary conditions at $S_L$ and $S_R$, respectively. At
$t\rightarrow -\infty$, $\Delta V^L = \Delta V^R = 0$, and $\Delta
V^L(t)$ and $\Delta V^R(t)$ are turned on near $t = 0$. We need
thus integrate Eq.~(\ref{e4rdm}) together with a Poisson equation
for the Coulomb potential inside the device region $D$ subject to
the boundary condition determined by the potentials at $S_{L}$ and
$S_{R}$. It is important to point out that $\iqa[t; \irhod]$ is
actually a nearly local quantity of the reduced system through the
local coupling matrix terms $h_{D\alpha}$ ($\alpha = L$ or $R$).
In this sense, our formalism for open electronic systems is not in
conflict with the "nearsightedness" concept of Kohn~\cite{near}.

\section{Two approximate schemes for self-energy density functionals}

\subsection{Complete second order approximation for dissipative
functional \label{cso}}

Eqs.~(\ref{e4qt}) and (\ref{e4qt2}) appear quite complicated. To
have an unambiguous interpretation of the dissipation term $\iqa$,
we further assume the KS Fock matrix $h_D$ is time-independent and
treat $G^<_D(t, \tau)$ by means of CS-QDT~\cite{qdt}.
Eq.~(\ref{e4qt}) is thus simplified to be (see
Appendix~\ref{app-iqa2} for details) \beq \label{q-phys}
   \iqa(t) = i \left\{ [\tilde{\Sigma}^>_\alpha(h_D), \sigma_D]^\dag
   + [\tilde{\Sigma}^<_\alpha(h_D), \bar{\sigma}_D]^\dag
   \right\},
\enq where $\bar{\sigma}_D \equiv I - \sigma_D$ is the reduced
single-hole density matrix of the reduced system. On the RHS of
Eq.~(\ref{q-phys}) a new commutator has been introduced for
arbitrary operators $A$ and $B$: \beq
   [A, B]^\dag \equiv AB - B^\dag A^\dag. \label{newcom}
\enq $\tslga$ are the causality-transformed counterparts of
$\Sigma^{<,>}_\alpha$, with $\Sigma^{<,>}_\alpha(t, \tau) =
\Sigma^{<,>}_\alpha(t - \tau)$ presumed, \emph{i.e.}, \be
   \tslga(h_D) & \equiv & \int_0^\infty dt\, \mbox{e}^{i h_D t}\,
   \Sigma^{<,>}_\alpha(t) \nonumber \\
   &=& \mp \, \Gamma^{(\pm)}_\alpha(h_D) \pm i
   \Lambda^{(\pm)}_\alpha(h_D), \label{tslga-def}
\en where $\Gamma^{(\pm)}_\alpha(h_D)$ and
$\Lambda^{(\pm)}_\alpha(h_D)$ are real symmetric matrices, and
associated with each other via the Kramers-Kronig
relation~\cite{qdt}. Therefore, Eq.~(\ref{q-phys}) can be expanded
as \be \label{q-phys2}
   \iqa(t) &=& i\left[\Gamma^{(-)}_\alpha(h_D), \sigma_D\right] +
   \left\{\Lambda^{(-)}_\alpha(h_D), \sigma_D\right\} - \nonumber \\
   && i\left[\Gamma^{(+)}_\alpha(h_D), \bar{\sigma}_D\right] -
   \left\{\Lambda^{(+)}_\alpha(h_D), \bar{\sigma}_D\right\}.
\en The physical meaning of Eq.~(\ref{q-phys2}) is clear and
intuitive: the first and third terms on its RHS account for the
energy shifting of occupied and virtual orbitals of the reduced
system due to the couplings with the lead $\alpha$, respectively;
and the second and fourth terms on its RHS are responsible for the
level broadening of occupied and virtual orbitals in $D$ due to
the lead $\alpha$ while contributing to the transient current,
respectively. The second term accounts for the electrons
leaving the device region, and the third term describes that
the holes hop onto the electrodes or the electrons enter
the device region from the electrodes.

\subsection{Solution for transient current with WBL approximation
and test on a model system \label{schemes}}

To simplify the solutions of
Eqs.~(\ref{eom4gretard})$-$(\ref{eom4gless}), the WBL
approximation~\cite{prb94win, pulse2c} may be adopted besides the
approximate $\ict$ operation (\emph{cf.} Eq.~(\ref{ct-approx})),
which involves the following assumptions for the leads: (i) their
band-widths are assumed to be infinitely large, such that the
summation over all the single-electron states in the leads can be
replaced by an integration over the entire energy range,
\emph{i.e.}, $\sum_{k\in\alpha} \rightarrow \int_{-\infty}^\infty
d\epsilon\, \eta_\alpha(\epsilon)$, (ii) their line-widths,
$\Lambda^{\alpha}_{k}(t,\tau)$, defined by the DOS at $S_{L}$ or
$S_{R}$ times the system-bath couplings, \emph{i.e.},
$\Lambda^{\alpha}_{k}(t,\tau) \equiv \pi\,\eta_{\alpha}
(\epsilon^{\alpha}_{k})\,\Gamma^{k_{\alpha}}(t,\tau)$, are treated
as energy independent, \emph{i.e.}, $\Lambda^{\alpha}_{k}(t,\tau)
\approx \Lambda^{\alpha}(t,\tau) \approx \Lambda^\alpha$, and
(iii) the level shifts of $L$ or $R$ are taken as a constant for
all energy levels, \emph{i.e.}, $\Delta\epsilon^{\alpha}_{k}(t)
\approx \Delta\epsilon^{\alpha}(t) = -\Delta V^{\alpha}(t)$, where
$\Delta V^{\alpha}(t)$ are the bias voltages applied on $L$ or $R$
at time $t$.

Within the WBL approximation, the self-energy functionals can be
expressed by~\cite{openprb}
\begin{eqnarray}
    \Sigma^{a}_{\alpha,nm}(\tau,t) &=& i\delta(t-\tau)\Lambda^{\alpha}_{nm}[\rho_D],
    \label{saa-wbl} \\
    \Sigma^{<}_{\alpha,nm}(\tau,t) &=&
    \frac{2i}{\pi}\,\exp\left\{i\int^{\tau}_{t}\Delta V^{\alpha}
    (\bar{t})\,d\bar{t}\right\} \,\Lambda^{\alpha}_{nm}[\rho_D]
    \nonumber\\
    && \times\,\left[\int_{-\infty}^{+\infty}f^{\alpha}(\epsilon)\,\mbox{e}^
    {\,i\epsilon(t-\tau)}d\epsilon\right]. \label{sla-wbl}
\end{eqnarray}
Here $\Delta V^\alpha(\bar{t})$ is the bias voltage applied on the
lead $\alpha$, and $\ila[\rho_D]$ is the line-width matrix due to
lead $\alpha$~\cite{openprb},
\begin{eqnarray}
   \ila_{nm}[\rho_D] &=& \pi\,\eta_\alpha(\epsilon_f)\left\langle
   h_{nk_f}\left[\rho_D, \rho_D(\br + N\mathbf{R})\right]\right. \nonumber   \\
   && \times\left.h_{k_f m}\left[\rho_D,
   \rho_D(\br + N\mathbf{R})\right]\right\rangle, \label{lambda-alpha}
\end{eqnarray}
where $\eta_\alpha(\epsilon_f)$ is the density of states for
$\alpha$ at its Fermi energy $\epsilon_f$, $k_f$ is a surface
state of $\alpha$ at $\epsilon_f$, and $\langle\cdots\rangle$
denotes the average over all surface states at $\epsilon_f$.
Eqs.~(\ref{saa-wbl})$-$(\ref{lambda-alpha}) provide thus the
explicit dependence of $\saa$ and $\sla$ on $\irhod$.

Note that $\sla[\rho_D]$ depends on the applied voltage $\Delta
V^\alpha(t)$ explicitly. In principle $\Delta V^\alpha(t)$ is a
functional of $\irhod$ as well. $\irhod$ is unknown and needs to
be solved. The potential $v(\br)$ in DFT formalism, which includes
the potentials from nuclei and external sources, is a functional
of electron density $\rho(\br)$. In any practical implementation
of DFT, $v(\br)$ is given and used to solve for $\rho(\br)$,
instead of determining $v(\br)$ from $\rho(\br)$. In our formalism
$\Delta V^\alpha(t)$ is given as a known function and used to
determine $\irhod$ in the same fashion.

Based on Eqs.~(\ref{saa-wbl})$-$(\ref{lambda-alpha}), the
dissipation term within the WBL approximation, $\iqawbl$, can be
obtained readily as follows (see Appendix~\ref{EOM4WBL} for
detailed derivations),
\begin{equation}\label{efinal}
    \iqawbl(t) = K^{\alpha}(t)
    +\left\{\ila[\rho_D], \sigma_D \right\}.
\end{equation}
Here the curly bracket on the RHS denotes an anticommutator, and
$K^{\alpha}(t)$ is a Hermitian matrix,
\begin{eqnarray}\label{k-wbl-1}
   K^{\alpha}(t) &=& -\frac{2i}{\pi}\,\bigg\{ \,U^{\alpha}(t)
   \int_{-\infty}^{\mu^{0}}\frac{d\epsilon\,
   \mbox{e}^{i\epsilon t}}{\epsilon-h_{D}(0)+i\,\Lambda} \nonumber
   \\
   && + \int_{-\infty}^{\mu^{0}}\left[ I - U^{\alpha}(t)\,
   \mbox{e}^{i\epsilon t} \right] \times \nonumber \\
   && \frac{d\epsilon }{\epsilon-h_{D}(t)+ i\Lambda
   + \Delta\epsilon^{\alpha}(t)}
   \bigg\}\Lambda^{\alpha} + \mbox{H.c.},
\end{eqnarray}
where $\mu^0$ is the chemical potential of the entire system, the
overall line-width $\Lambda = \sum_\alpha \ila$, and the effective
propagator of the reduced system $U^\alpha(t)$ is
\begin{eqnarray}\label{u-alpha}
   U^{\alpha}(t) &=& \mbox{e}^{-i\int_0^t \left[h_D(\tau) -i\Lambda
   -\dea(\tau)\right]d\tau}.
\end{eqnarray}

\begin{figure}
\includegraphics[scale=0.35]{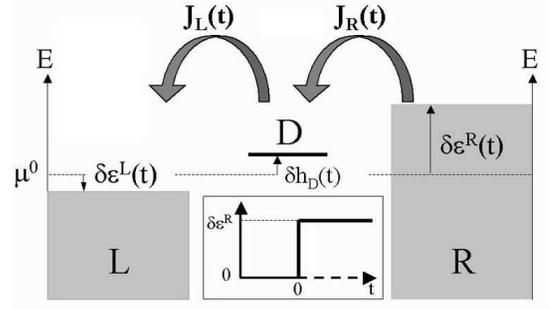}
\caption{\label{onesite} Model system for the test of the WBL
self-energy functionals where a single site spans the device
region $D$. Transient currents through leads $L$ and $R$, $J_L(t)$
and $J_R(t)$, are simulated. The inset shows the time-dependent
level shift of lead $R$. }
\end{figure}

\begin{figure}
\includegraphics[scale=0.55]{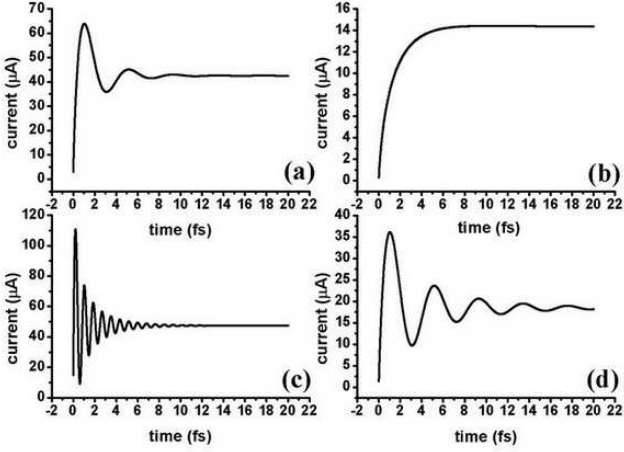}
\caption{\label{wbl-awbl} The calculated transient current through
$S_R$ within the WBL scheme. We set $\mu^0 = h_{D}(0) = 0$ for the
ground state; and $\Delta\epsilon^L(t) = 0$ and
$\Delta\epsilon^R(t) = \Delta\epsilon^R\, (1 - \mbox{e}^{-t / a})$
after switch-on. The above panels show different cases where (a)
$\Delta\epsilon^R = 2$ eV, $\Lambda^L = \Lambda^R = 0.1$ eV; (b)
$\Delta\epsilon^R = 0.2$ eV, $\Lambda^L = \Lambda^R = 0.1$ eV; (c)
$\Delta\epsilon^R = 10$ eV, $\Lambda^L = \Lambda^R = 0.1$ eV; and
(d) $\Delta\epsilon^R = 2$ eV, $\Lambda^L = \Lambda^R = 0.04$ eV,
respectively. }
\end{figure}

\subsection{Numerical test of wide-band limit approximation}

The WBL approximated self-energy functionals are then tested by
calculations on a model system which has previously been
investigated by Maciejko, Wang and Guo~\cite{pulse2c}. In this
model system the device region $D$ consists of a single site
spanned by only one atomic orbital (see Fig.~\ref{onesite}). Exact
transient current driven by a step voltage pulse has been obtained
from NEGF simulations~\cite{pulse2c}, and the authors concluded
that the WBL approximation yields reasonable results provided that
the band-widths of the leads are five times or larger than the
coupling strength between $D$ and $L$ or $R$. The computational
details are as follows. The entire system ($L$ + $R$ + $D$) is
initially in its ground state with the chemical potential $\mu^0$.
External bias voltages are switched on from the time $t = 0$,
which results in transient current flows through the leads $L$ and
$R$. $\delta h_D(t) \equiv h_D(t) - h_D(0)$, $\del(t)$ and
$\der(t)$ are the level shifts of $D$, $L$ and $R$ at time $t$,
respectively. In our works we take $\delta h_D(t) = \frac{1}{2}
\left[\del(t) + \der(t)\right]$, $\Delta\epsilon^L(t) = 0$, and
$\Delta\epsilon^R(t) = \Delta\epsilon^R\, (1 - \mbox{e}^{-t /
a})$, where $a$ is a positive constant. The real analytic level
shift $\Delta \epsilon^R(t)$ resembles perfectly a step pulse as
$a\rightarrow 0^{+}$ (see the inset of Fig.~\ref{onesite}). The
calculation results are demonstrated in Fig.~\ref{wbl-awbl}. We
choose exactly the same parameter set as that adopted for Fig.~2
in Ref.~\cite{pulse2c}, and the resulting transient current,
represented by Fig.~\ref{wbl-awbl}(a), excellently reproduces the
WBL result in Ref.~\cite{pulse2c}, although the numerical
procedures employed are distinctively different. The comparison
confirms evidently the accuracy of our formalism. From
Fig.~\ref{wbl-awbl}(a)$-$(c) it is observed that with the same
line-widths $\Lambda^{\alpha}$, a larger level shift
$\Delta\epsilon^R$ results in a more fluctuating current, whereas
by comparing (a) and (d) we see that under the same
$\Delta\epsilon^R$, the current decays more rapidly to the steady
state value with the larger $\Lambda^{\alpha}$.

By transforming its integrand into a diagonal representation, the
integration over energy in Eq.~(\ref{k-wbl-1}) can be carried out
readily. Therefore, $\iqawbl$ are evaluated straightforwardly,
which makes the above solution procedures for transient dynamics
within the WBL approximation a practical routine for subsequent
TDDFT calculations.

\section{TDDFT calculations of transient current through molecular
devices \label{tddft-wbl}}

\subsection{Numerical procedures \label{proc}}

With the EOM~(\ref{e4rdm}) and the WBL approximation for the
self-energy functionals $\saa[\rho_D]$ and $\sla[\rho_D]$, it is
now straightforward to investigate the transient dynamics of open
electronic systems. All our first-principles calculations are
carried out with a self-developed package
\texttt{LODESTAR}~\cite{lodestar}.

The ground state properties of the reduced system at $t = 0$ are
determined by following the partitioned scheme approach adopted in
conventional DFT-NEGF method~\cite{jcpratner, prbguo, prbywt,
transiesta}. Different from the popular
periodic-boundary-condition-based approach~\cite{transiesta,
prbywt, seqquest0}, what we employ is a molecular-cluster-based
technique~\cite{lodestar}. The ground state KS Fock matrix of an
extended cluster, covering not only the device region $D$ but also
portions of leads $L$ and $R$, is calculated self-consistently by
conventional DFT method with local density approximation (LDA) for
the XC functional~\cite{ks}. Its diagonal blocks corresponding to
the leads $L$ and $R$ are then extracted and utilized to evaluate
the surface Green's function of isolated lead $\alpha$ ($L$ or
$R$), $g^r_\alpha = g^r_\alpha[\mu^0; \irhoa^{\ict}[\rho_D]]$, by
applying the translational invariance~\cite{surfg} (\emph{cf.}
Eq.~(\ref{ct-approx})). In this way the possible misalignment for
the chemical potentials of the isolated leads $L$ and $R$,
especially when they are made of different materials, can be
avoided so long as the extended cluster is chosen large enough. In
an orthogonal atomic orbital basis set, the line-widths
$\ila[\rho_D]$ within the WBL approximation are obtained from
$g^r_\alpha$ via \beq \label{lambda-gs}
   \ila[\rho_D] = -\Im\left\{ h_{D\alpha}\, g^r_\alpha\left[\mu^0;
   \irhoa^{\ict}[\rho_D]\right]\, h_{\alpha D} \right\}. \enq
At $t = 0$ the left-hand side (LHS) of the Eq.~(\ref{e4rdm})
vanishes. The EOM~(\ref{e4rdm}) reduces thus to a nonlinear
equation for $\sigma_D(0)$, and can be solved readily by employing
the NEGF approach as follows, \beq \label{sigma-gs}
  \sigma_D(0) = \frac{2}{\pi} \int_{-\infty}^{\mu^0} d\epsilon
  \, G^{r,0}_D(\epsilon)\, \Lambda\,  G^{a,0}_D(\epsilon),
\enq where \beq \label{gr-gs}
  G^{r,0}_D(\epsilon) = [G^{a,0}_D(\epsilon)]^\dag =
  \left[ \epsilon - h_D(0) + i\Lambda \right]^{-1}.
\enq Eq.~(\ref{sigma-gs}) provides the initial condition for the
EOM~(\ref{e4rdm}).

The molecular device is switched on by a step-like voltage $\Delta
V^R(t) = -\Delta\epsilon^R(t) = \Delta V^R (1 - \mbox{e}^{-t /
a})$ applied on the right lead with $a \rightarrow 0^+$ (see the
inset of Fig.~\ref{onesite}), while $\Delta V^L(t) = 0$. The
self-energy functionals $\saa[\rho_D]$ and $\sla[\rho_D]$ can be
evaluated through Eqs.~(\ref{saa-wbl})$-$(\ref{sla-wbl}) and
(\ref{lambda-gs}). The dynamic response of the reduced system is
obtained by solving the EOM~(\ref{e4rdm}) in time domain within
the adiabatic LDA (ALDA)~\cite{casida} for the XC functional. The
induced KS Fock matrix of the reduced system, $\delta h_D(t)
\equiv h_D(t) - h_D(0)$, is comprised of Hartree and XC
components~\cite{ldmtddft}, \emph{i.e.}, \beq
  \delta h_D(t) = \delta h^H_D(t) + \delta h^{XC}_D(t),
\enq where \beq
  \delta h^H_{ij}(t) = \int_D d\br\, \phi^\ast_i(\br)\,
  \delta v^H(\br, t)\, \phi_j(\br).
\enq Here the Hartree potential $\delta v^H(\br, t)$ satisfies the
following Poisson equation for the device region $D$ subject to
boundary conditions $\Delta V^{\alpha}(t)$ at every time $t$:
\begin{equation}\label{v-hart}
\left\{
 \begin{array}{rcl}
    \nabla^{2}\,\delta v^H(\br, t) &=&
    -4\pi\,\delta \irhod \\
    \left. \delta v^H(\br, t)\right\vert_{S_L} &=& \dvl(t) \\
    \left. \delta v^H(\br, t)\right\vert_{S_R} &=& \dvr(t) .
 \end{array}
 \right.
\end{equation}
To save computational resources we calculate $\delta h^{XC}_D(t)$
to its first-order change due to the switch-on potential:
\begin{eqnarray}
   \delta h^{XC}_{ij}(t) &=& \sum_{mn\in D}V^{XC}_{ijmn}\,
   \left[\sigma_{mn}(t) - \sigma_{mn}(0)\right],\label{xcksf}\\
V^{XC}_{ijmn} &=& \int_D d\mathbf{r}\,\phi^{\ast}_m(\mathbf{r})
\phi_n(\mathbf{r})\frac{\delta v^{XC}[\br, t; \rho_D]}
{\delta\rho_D(\mathbf{r},t)} \nonumber \\
&& \times\,\phi^{\ast}_i(\mathbf{r}) \phi_j(\mathbf{r}),
\label{vxcijmn}
\end{eqnarray}
where $v^{XC}[\br, t; \rho_D]$ is the XC potential. The reduced
system is propagated from $t = 0$ following the EOM~(\ref{e4rdm})
by the fourth-order Runge-Kutta algorithm~\cite{kutta} with the
time step $0.02$ fs. Virtually the same results are yielded by
adopting a much smaller time step, which justifies the accuracy of
our time evolution scheme.

\subsection{Calculation on a graphene-alkene-graphene system \label{gra-alk-gra}}

\begin{figure}
\includegraphics[scale=0.45]{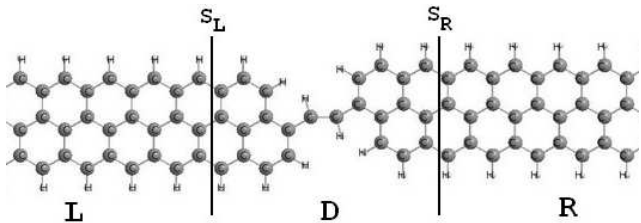}
\caption{\label{sys-gag} A graphene-alkene-graphene system adopted
in TDDFT calculations. }
\end{figure}

A realistic molecular device depicted in Fig.~\ref{sys-gag} is
taken as the open system under investigation. The device region
$D$ containing $24$ carbon and $12$ hydrogen atoms is spanned by
the 6-31 Gaussian basis set, \emph{i.e.}, altogether $240$ basis
functions for the reduced system. The leads are
quasi-one-dimensional graphene ribbons with dangling bonds
saturated by hydrogen atoms, and the entire system is on a same
plane. The extended cluster contains totally $134$ atoms.

\begin{figure}
\begin{center}
\includegraphics[scale=0.55]{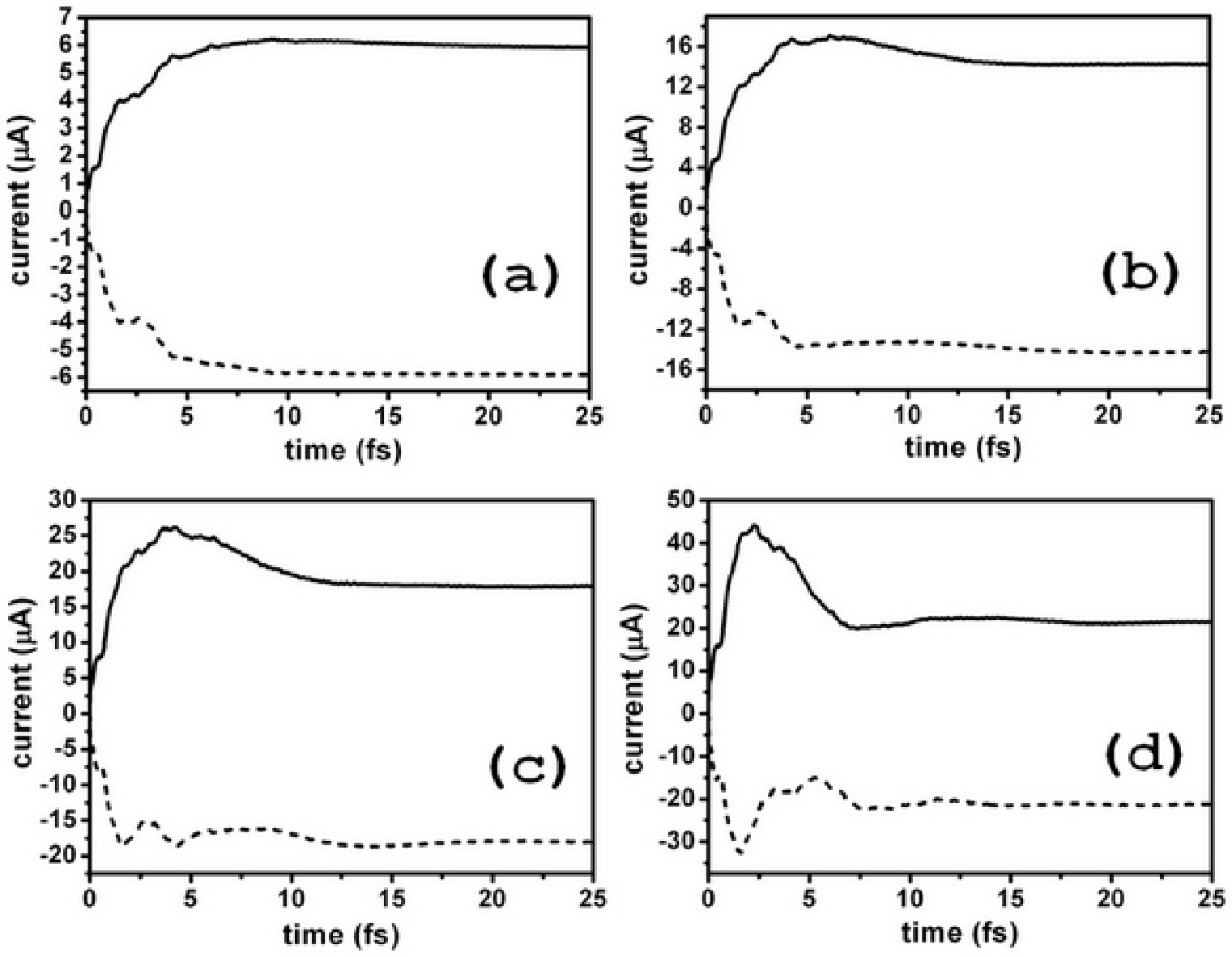}
\caption{\label{td-gag} The solid (dashed) curve represents the
transient current through the interface $S_R$ ($S_L$) of the
graphene-alkane-graphene system driven by a step-like voltage
applied on the lead $R$ with the amplitude (a) $\Delta V^R = -0.1$
V, (b) $\Delta V^R = -0.3$ V, (c) $\Delta V^R = -0.5$ V, and (d)
$\Delta V^R = -1.0$ V. }
\end{center}
\end{figure}

In Fig.~\ref{td-gag} we plot the calculated transient currents
through the interfaces $S_L$ and $S_R$, $J_L(t)$ and $J_R(t)$,
under various turn-on voltages. As depicted in Fig.~\ref{td-gag},
$J_L(t)$ and $J_R(t)$ increase rapidly during the first few fs and
then approach gradually towards their steady state values. This
agrees with previous investigations on model systems \cite{kurth1,
pulse2c}. The steady currents through $S_L$ and $S_R$ are (a)
$-5.9~\mu$A and $5.9~\mu$A, (b) $-14.2~\mu$A and $14.2~\mu$A, (c)
$-18.0~\mu$A and $18.0~\mu$A, and (d) $-21.3~\mu$A and
$21.3~\mu$A, respectively, and thus cancel each other out exactly,
as they should. By comparison of panels (a)$-$(d) it is obvious
that a larger turn-on voltage results in a more conspicuous
overshooting for the transient current. Complex fluctuations are
also observed for the time-dependent currents, which are due to
the various eigenvalues possessed by the nonnegative definite
line-widths $\Lambda^{\alpha}$ with their magnitudes ranging from
$0$ to $4.1$ eV, corresponding to various dissipative channels
between $D$ and $L$ or $R$. From Fig.~\ref{td-gag}, the
characteristic switch-on time for the graphene-alkene-graphene
system is estimated as about $10 \sim 15$ fs for applied bias
voltages ranging from $0.1$ V to $1.0$ V. For much higher turn-on
voltages the linearized form of $\delta h^{XC}_D(t)$
(Eq.~(\ref{xcksf})) becomes inadequate, which makes such a TDDFT
calculation computationally demanding with our present coding.

It is noted that the reduced system remains in its ground state in
absence of an applied bias voltage. This is confirmed by a free
propagation for the reduced system. During the course the
transient current $J_L(t)$ or $J_R(t)$ vanishes correctly at every
time $t > 0$. This thus validates that the WBL approximated
self-energy functionals derived from the partitioned scheme
(\emph{cf.} Eq.~(\ref{e4qt})) is well adapted to a TDDFT
formalism.

\subsection{Calculation on a CNT-alkene-CNT system \label{cnt-alk-cnt}}

\begin{figure}
\begin{center}
\hspace{-0.5cm}
\includegraphics[scale=0.5]{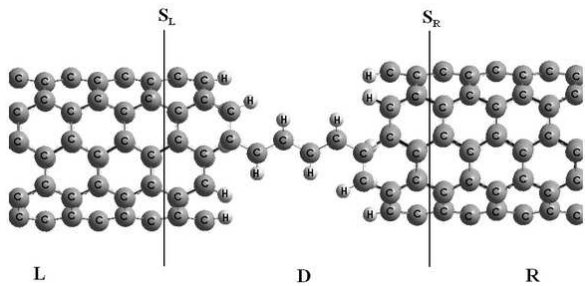}
\caption{\label{sys-cac} A CNT-alkene-CNT system adopted in TDDFT
calculations. }
\end{center}
\end{figure}

\begin{figure}
\hspace{-0.9cm}
\includegraphics[scale=0.9]{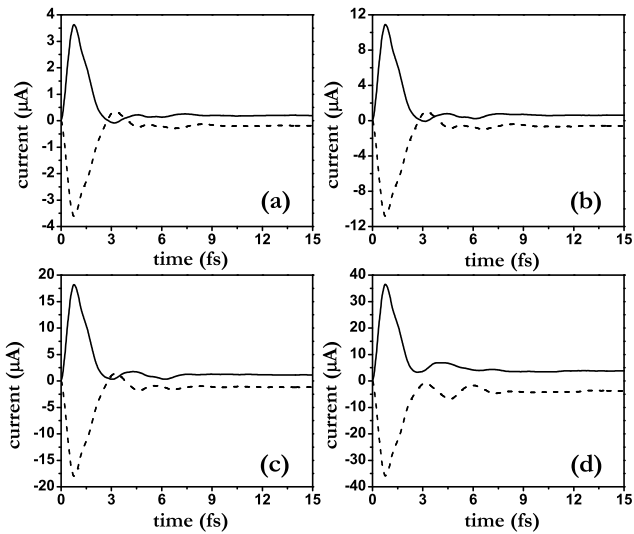}
\caption{\label{td-cac} The solid (dashed) curve represents the
transient current through the interface $S_R$ ($S_L$) of the
CNT-alkene-CNT system driven by a step-like voltage applied on the
lead $R$ with the amplitude (a) $\Delta V^R = -0.1$ V, (b) $\Delta
V^R = -0.3$ V, (c) $\Delta V^R = -0.5$ V, and (d) $\Delta V^R =
-1.0$ V. }
\end{figure}

The second molecular device we calculate is sketched in
Fig.~\ref{sys-cac}, where a linear alkene is connected to
semi-infinite single-walled carbon nanotubes (CNT) (5, 5) at its
both ends. The device region $D$ consists of 88 carbon and 22
hydrogen atoms, \emph{i.e.}, altogether 836 basis functions for
the reduced system. The extended cluster for the ground state
calculation contains totally 290 atoms. The calculated transient
currents driven by step-like turn-on voltages $\dvr(t)$ (see the
inset of Fig.~\ref{onesite}) are plotted in Fig.~\ref{td-cac}.
Here we have set $\dvl = 0$. The switch-on time for the
CNT-alkene-CNT system is about $10$ fs for applied voltages
ranging from $0.1$ V to $1.0$ V.

\subsection{Calculation on an Al-C7-Al system \label{al-c7-al}}

\begin{figure}
\begin{center}
\includegraphics[scale=0.5]{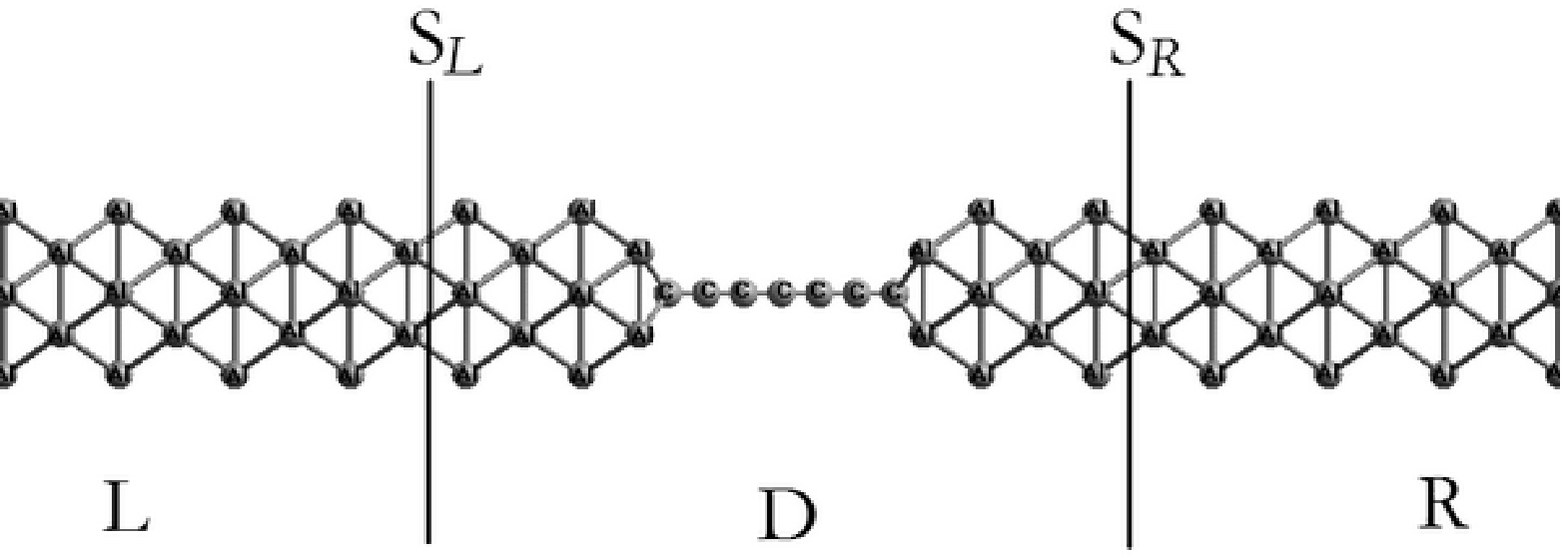}
\caption{\label{sys-aca} A linear carbon chain is sandwiched
between two Al leads in the (001) direction of bulk Al. }
%
\hspace{-1.0cm}
\includegraphics[scale=0.9]{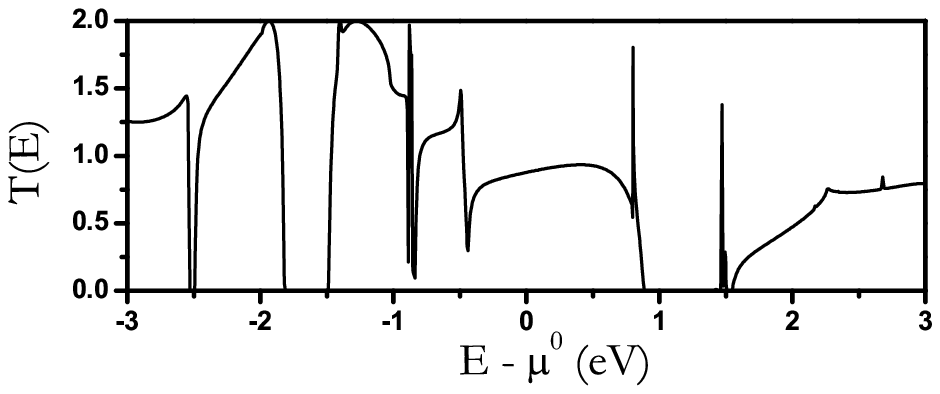}
\caption{\label{te-aca} Non-WBL transmission coefficient
$T(\epsilon; \dvr=0 \mbox{V})$ of the Al-C7-Al system.}
%
\hspace{-1.0cm}
\includegraphics[scale=0.9]{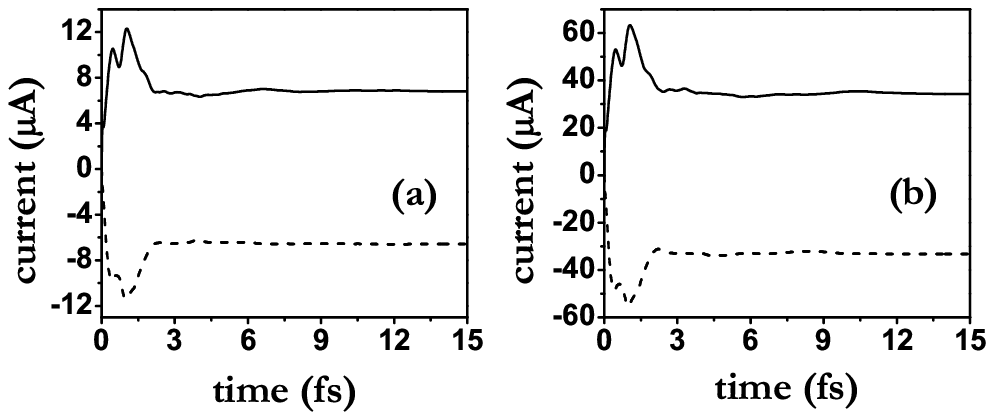}
\caption{\label{td-aca} The solid (dashed) curve represents the
transient current through the interface $S_R$ ($S_L$) of the
Al-C7-Al system driven by a step-like voltage applied on the lead
$R$ with the amplitude (a) $\Delta V^R = -0.1$ V, and (b) $\Delta
V^R = -0.5$ V. }
\end{center}
\end{figure}

Another open system adopted in our first-principles calculations
is depicted in Fig.~\ref{sys-aca}, where a linear chain of seven
carbon atoms is embedded between two semi-infinite Al leads in the
(001) direction of bulk Al. The current-voltage characteristics of
this Al-C7-Al system with the same geometric configuration has
been investigated extensively~\cite{prbywt, transiesta}. In our
calculation, the device region $D$ consists of 7 carbon and 18 Al
atoms, \emph{i.e.}, altogether 297 basis functions for the reduced
system, and the extended cluster for ground state calculation
contains totally 115 atoms.

The calculated non-WBL transmission coefficient, $T(\epsilon;
\dvr=0 \mbox{V})$, is plotted in Fig.~\ref{te-aca}. The main
features of our result agree reasonably with those exhibited in
literature~\cite{prbywt, transiesta}. The quantitative
discrepancies may be due to the different techniques employed. For
instance, a finite molecular cluster is explicitly treated in our
calculation, whereas an infinite periodic system is considered in
Refs~\cite{prbywt, transiesta}, and also the basis set and XC
functional adopted are distinctively different. The calculated
transient currents driven by step-like turn-on voltages $\dvr(t)$
(see the inset of Fig.~\ref{onesite}) are plotted in
Fig.~\ref{td-aca}. The switch-on time for the Al-C7-Al system is
about $3\sim5$ fs for applied voltages ranging from $0.1$ V to
$0.5$ V.

\section{Discussion and Summary \label{summary}}

Kurth \emph{et al.} have proposed a practical TDDFT approach
combined with the partition-free scheme~\cite{kurth1}. A number of
relevant technical issues have been addressed, for instance, how
the intractable propagation of the KS orbitals of an infinitely
large system is transformed into the time evolution of KS orbitals
in a finite open system subject to correct boundary conditions,
how the time-dependent KS equation for the entire system is
discretized in both $\br$ and $t$ spaces, \emph{etc.}. The
performance of their approach has been illustrated by calculations
for one-dimensional model systems. Our first-principles formalism
for open electronic systems is fundamentally different: (i) In our
method the KS reduced single-electron density matrix is used as
the basic variable while in Ref.~\cite{kurth1} the occupied KS
single-electron orbitals are propagated. (ii) The concept of
self-energy functional is introduced in our formalism. In
principle the self-energy functional depends only on the electron
density function of the reduced system, and hence we need only
focus on the reduced system of interest without treating
explicitly the environment. The influence of the environment
enters via boundary conditions and the self-energy functionals.
This is not only for quantum transport phenomena, but also for any
dynamic process in any open electronic system. In this sense we
expect the EOM~(\ref{e4rdm}) to be a general recipe for open
system problems. (iii) Our EOM is formally analogous to the master
equations derived from the conventional QDT~\cite{qdt}. From this
perspective, well-established methods and techniques of QDT may be
employed to improve the evaluations of self-energy functionals and
the dissipation term $\iqa[t; \irhod]$ systematically. For
instance, another EOM has recently been proposed by Cui~\emph{et
al.} based on the CS-QDT with a self-consistent Born approximation
(SCBA)~\cite{csqdt-scba}.

In conventional QDT~\cite{qdt} the key quantity is the reduced
system density matrix, whereas in Eq.~(\ref{e4rdm}) the basic
variable is the reduced single-electron density matrix, which
leads to the drastic reduction of the degrees of freedom in
numerical simulation. Linear-scaling methods such as the
localized-density-matrix (LDM) method~\cite{ldmtddft, ldm} may
thus be adopted to further speed up the solution process of
Eq.~(\ref{e4rdm}). Therefore, Eq.~(\ref{e4rdm}) provides an
accurate and convenient formalism to investigate the dynamic
properties of open systems.

It is worth mentioning that our first-principles method for open
systems applies to the same phenomena, properties or systems as
those intended by Hohenberg and Kohn~\cite{hk}, Kohn and
Sham~\cite{ks}, and Runge and Gross~\cite{tddft}, {\it i.e.},
where the exchange-correlation energy is a functional of electron
density only, $E_{XC} = E_{XC}[\rho(\mathbf{r})]$. This is true
when the interaction between the electric current and magnetic
field is negligible. However, in the presence of a strong magnetic
field, $E_{XC} =
E_{XC}[\rho(\mathbf{r}),\mathbf{j}_{p}(\mathbf{r})]$ or $E_{XC} =
E_{XC}[\rho(\mathbf{r}),\mathbf{B}(\mathbf{r})]$, where
$\mathbf{j}_p(\mathbf{r})$ is the paramagnetic current density and
$\mathbf{B}(\mathbf{r})$ is the magnetic field~\cite{magdft}. In
such a case, our first-principles formalism needs to be
generalized to include $\mathbf{j}_{p}(\mathbf{r})$ or
$\mathbf{B}(\mathbf{r})$. Of course, $\mathbf{j}_{p}(\mathbf{r})$
or $\mathbf{B}(\mathbf{r})$ should be an analytical function in
space.

To summarize, we have proven the existence of a first-principles
method for time-dependent open electronic systems, and developed a
formally closed TDDFT formalism by introducing the concept of
self-energy functionals. In principle the self-energy functionals
depend only on the electron density function of the reduced
system. With an efficient WBL approximation for self-energy
functionals, we have applied the first-principles formalism to
carry out TDDFT calculations for transient current through
realistic molecular devices. This work greatly extends the realm
of density-functional theory.

\begin{acknowledgements}
Authors would thank Hong Guo, Shubin Liu, Jiang-Hua Lu, Zhigang
Shuai, K. M. Tsang, Bing Wang, Jian Wang, Arieh Warshel, Yijing
Yan and Weitao Yang for stimulating discussions. Support from the
Hong Kong Research Grant Council (HKU 7010/03P) is gratefully
acknowledged.
\end{acknowledgements}

\appendix

\section{Derivation of Eq.~(\ref{e4qt}) with the Keldysh
formalism \label{derivee4qt}}

\begin{figure}[h]
\hspace{1.0cm}
\includegraphics[scale=0.42]{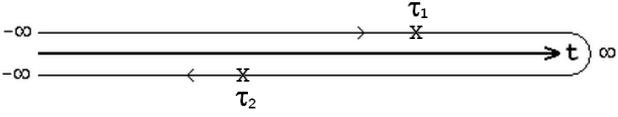}
\caption{\label{kcontour} The Keldysh time contour on which
nonequilibrium Green's function theory is constructed. On the
contour, the time $\tau_1$ is earlier than $\tau_2$ even though
its real-time projection appears larger. }
\end{figure}

In the Keldysh formalism \cite{keldysh}, the nonequilibrium
single-electron Green's function $G_{k_{\alpha},m}(t,t')$ is
defined by
\begin{equation}\label{contourg0}
    G_{k_{\alpha}m}(t,t') \equiv -i\left\langle T_{C}\!\left
    \{a_{k_{\alpha}}\!(t)\,a_{m}^{\dag}(t')\right\}\right\rangle,
\end{equation}
where $T_{C}$ is the contour-ordering operator along the Keldysh
contour \cite{prb94win, keldysh} (see Fig.~\ref{kcontour}). Its
lesser component, $G^{<}_ {k_{\alpha},m}(t,t')$, is defined by
\begin{equation}\label{glesser0}
    G^{<}_{k_{\alpha}m}(t,t') \equiv i\langle\,a^{\dag}_{m}(t')\,
    a_{k_{\alpha}}(t)\rangle.
\end{equation}
The formal NEGF theory has exactly the same structure as that of
the time-ordered Green's function at zero temperature
\cite{prb94win, mahan1}. Thus, the Dyson equation for
$G_{k_{\alpha}m}(t,t')$ can be written as
\begin{equation}\label{eom4gc}
    G_{k_{\alpha}m}(t,t')=\sum_{l\in D}\int_{C}d\tau\,g_{k_{\alpha}}
    (t,\tau)\,h_{k_{\alpha}l}(\tau)\,G_{lm}(\tau,t'),
\end{equation}
where $G_{lm}(\tau,t')$ and $g_{k_{\alpha}}(t,\tau)$ are the
contour-ordered Green's functions for the reduced system $D$ and
the isolated semi-infinite lead $\alpha$ ($L$ or $R$),
respectively, and the integration over $\tau$ on the RHS is
performed along the entire Keldysh contour (see Fig.
\ref{kcontour}).

\vspace{0.2cm}

$G^r_{lm}(\tau, t')$, $G^a_{lm}(\tau, t')$ and $G^<_{lm}(\tau,
t')$ denote the retarded, advanced and lesser components of
$G_{lm}(\tau,t')$, respectively. Their definitions are as follows,
\begin{eqnarray}
   G^r_{lm}(\tau, t') &\equiv& -i \vartheta(\tau - t')\langle \left\{
   a_l(\tau), a^{\dag}_m(t') \right\} \rangle, \label{gr-lm} \\
   G^a_{lm}(\tau, t') &\equiv& i  \vartheta(t' - \tau)\langle \left\{
   a_l(\tau), a^{\dag}_m(t') \right\} \rangle, \label{ga-lm} \\
   G^<_{lm}(\tau, t') &\equiv& i \langle a^{\dag}_m(t') a_l(\tau)
   \rangle, \label{gless-lm}
\end{eqnarray}
where $\vartheta(\tau - t')$ is the Heaviside step function, and
the expectation values $\langle\cdots\rangle$ are taken at the
ground state of the entire system at $t = -\infty$, \emph{i.e.},
when the reduced system and the environment are completely
decoupled. $G^r_{lm}(\tau, t')$ and $G^<_{lm}(\tau, t')$ are to be
calculated via their EOMs (\ref{eom4gretard})$-$(\ref{eom4gless}).
The related self-energies $\Sigma^a_\alpha(t, \tau)$ and
$\Sigma^<_\alpha(t, \tau)$ are evaluated through \be
\label{selfes}
    \Sigma^a_{\alpha, ln}(t, \tau) & = & \sum_{\ka\in\alpha}
    h_{lk_{\alpha}}(t)\,g^{a}_{k_{\alpha}}(t, \tau)
    \,h_{k_{\alpha}n}(\tau), \\
    \Sigma^<_{\alpha, ln}(t, \tau) & = & \sum_{\ka\in\alpha}
    h_{lk_{\alpha}}(t)\,g^{<}_{k_{\alpha}}(t, \tau)
    \,h_{k_{\alpha}n}(\tau),
\en
for $\alpha = L$ or $R$. Here $\gak$ and $\glk$ are the advanced
and lesser surface Green's functions for the isolated lead
$\alpha$ ($L$ or $R$)~\cite{prb94win}.

Applying the analytical continuation rules of Langreth
\cite{langreth}, we have
\begin{eqnarray}\label{glesser-mk-detail}
    G^{<}_{mk_{\alpha}}(t',t) &\equiv& i\langle a^{\dag}_{k_{\alpha}}
    (t)\,a_{m}(t')\rangle \nonumber \\
    &=&  -\left[ G^{<}_{k_{\alpha}m}(t,t') \right]^{\ast}
    \nonumber \\
    &=&\sum_{l\in D}\int_{-\infty}^{\infty}
    d\tau\,h_{lk_{\alpha}}(\tau)\Big[g^{<}_{k_{\alpha}}(\tau,t)\,
    G^{r}_{ml}(t',\tau) \nonumber \\
    && +\,\, g^{a}_{k_{\alpha}}(\tau,t)\,G^{<}_{ml}(t',\tau)\Big]
\end{eqnarray}
by adopting the following equalities:
\begin{eqnarray}\label{equalities0}
    G^{r}_{ml}(t',\tau) &=& \left[G^{a}_{lm}(\tau,t')
    \right]^{\ast}, \nonumber \\
    G^{<}_{ml}(t',\tau) &=& -\left[G^{<}_{lm}(\tau,t')
    \right]^{\ast}, \nonumber \\
    g^{a}_{k_{\alpha}}(\tau,t) &=& \left[g^{r}_{k_{\alpha}}
    (t,\tau)\right]^{\ast}, \nonumber \\
    g^{<}_{k_{\alpha}}(\tau,t) &=& -\left[g^{<}_{k_{\alpha}}
    (t,\tau)\right]^{\ast}.
\end{eqnarray}
Note that $\sigma_{m k_{\alpha}}(t)$ is precisely the lesser
Green's function of identical time variables, \emph{i.e.}, \beq
 \sigma_{m k_{\alpha}}(t) = -i\left.G^{<}_{m \ka}(t,t')\right\vert_
{t'=t}. \label{glttau} \enq By inserting
Eqs.~(\ref{glesser-mk-detail}) and (\ref{glttau}) into
Eq.~(\ref{qterm0}), Eq.~(\ref{e4qt}) can be recovered
straightforwardly.

\section{The dissipation term
$\mathbf{Q}_{\mathbf{\alpha}}$ in partition-free and partitioned
schemes \label{app-iqa}}

For brevity, $\sum_{\alpha=L,R}\sum_{k\in\alpha}$ will be
shortened to $\sum_{\ka}$. The Hamiltonian of the entire
noninteracting KS system is \be
  H(t) &=& \sum_{mn\in D}h_{mn}(t)a^\dag_m a_n + \sum_{\ka}
  \epsilon_{\ka}(t) a^\dag_{\ka} a_{\ka} \nonumber \\
  && + \sum_{m\in D}\sum_{\ka}
  \left[h_{m\ka}(t)a^\dag_m a_{\ka} + \mbox{H.c.}\right]. \label{fock-all}
\en  Initially (at $t = t_0$) the entire KS system is in its
ground state $\Psi(t_0)$ (denoted by $\vert 0 \rangle$ hereafter),
\emph{i.e.}, $H(t_0)\Psi(t_0) = E_0 \Psi(t_0)$. We define the
following Heisenberg creation and annihilation operators ($\hbar
\equiv 1$): \be
  a^\dag_m(t) &\equiv& \mbox{e}^{i\int_{t_0}^t H(\tau)d\tau} a^\dag_m \mbox{e}^{-i\int_{t_0}^t
  H(\tau)d\tau}, \nonumber \\
  a_m(t) &\equiv& \mbox{e}^{i\int_{t_0}^t H(\tau)d\tau} a_m \mbox{e}^{-i\int_{t_0}^t
  H(\tau)d\tau}, \nonumber \\
  a^\dag_{\ka}(t) &\equiv& \mbox{e}^{i\int_{t_0}^t H(\tau)d\tau} a^\dag_{\ka} \mbox{e}^{-i\int_{t_0}^t
  H(\tau)d\tau}, \nonumber \\
  a_{\ka}(t) &\equiv& \mbox{e}^{i\int_{t_0}^t H(\tau)d\tau} a_{\ka} \mbox{e}^{-i\int_{t_0}^t
  H(\tau)d\tau}, \label{a-adag}
\en which satisfy their respective EOMs ($\partial_t \equiv
\frac{\partial}{\partial t}$): \be
  \partial_t a^\dag_m(t) &=& i\sum_{i\in D}a^\dag_i(t)h_{im}(t) +
  i\sum_{\ka}a^\dag_{\ka}(t)h_{\ka m}(t), \nonumber \\
  \partial_t a_m(t) &=& -i\sum_{i\in D}h_{mi}(t)a_i(t) -
  i\sum_{\ka}h_{m\ka}(t)a_{\ka}(t), \nonumber \\
  \partial_t a^\dag_{\ka}(t) &=& i\sum_{i\in D}a^\dag_i(t)h_{i\ka}(t) +
  i\,\epsilon_{\ka}(t)a^\dag_{\ka}(t), \nonumber \\
  \partial_t a_{\ka}(t) &=& -i\sum_{i\in D}h_{\ka i}(t)a_i(t)
  -i\,\epsilon_{\ka}(t)a_{\ka}(t),
\en with the initial conditions: $a^\dag_m(t_0) = a^\dag_m$,
$a_m(t_0) = a_m$, $a^\dag_{\ka}(t_0) = a^\dag_{\ka}$, and
$a_{\ka}(t_0) = a_{\ka}$.

The retarded, advanced and lesser surface Green's functions for
the isolated lead $\alpha$ ($L$ or $R$) are defined as follows,
\be
   g^{r,a}_{\ka}(t,\tau) &\equiv& \mp i \vartheta(\pm t \mp \tau)\langle \alpha \vert
   \{ b_{\ka}(t), b^\dag_{\ka}(\tau) \} \vert \alpha \rangle,
   \label{graka} \\
   g^<_{\ka}(t,\tau) &\equiv& i \langle \alpha \vert
   b^\dag_{\ka}(\tau)b_{\ka}(t) \vert \alpha \rangle, \label{glka}
\en where the curly bracket on the RHS of Eq.~(\ref{graka})
denotes an anticommutator, and $\vert\alpha\rangle$ is the ground
state wavefunction corresponding to the initial lead Hamiltonian
$H_\alpha(t_0)$: \beq
   H_\alpha(t) = \sum_{k\in\alpha}
   \epsilon_{\ka}(t)a^\dag_{\ka}a_{\ka}.
\enq The Heisenberg operators in Eqs.~(\ref{graka}) and
(\ref{glka}) are defined by \be
  b^\dag_{\ka}(t) &\equiv& \mbox{e}^{i\int_{t_0}^t H_\alpha(\tau)d\tau} a^\dag_{\ka} \mbox{e}^{-i\int_{t_0}^t
  H_\alpha(\tau)d\tau}, \nonumber \\
  b_{\ka}(t) &\equiv& \mbox{e}^{i\int_{t_0}^t H_\alpha(\tau)d\tau} a_{\ka} \mbox{e}^{-i\int_{t_0}^t
  H_\alpha(\tau)d\tau}.
\en

We then define the retarded, advanced and lesser Green's functions
of the entire KS system via their matrix elements as follows, \be
 G^{r,a}_{ij}(t, \tau) &\equiv& \mp i\vartheta(\pm t \mp \tau)\langle 0 \vert \{
 a_i(t), a^\dag_j(\tau) \} \vert 0 \rangle, \\
  G^<_{\ka j}(t, \tau) &\equiv& i\langle 0 \vert a^\dag_{j}(\tau)
  a_{\ka}(t)\vert 0 \rangle, \label{glkj} \\
  G^<_{\ka \pb}(t, \tau) &\equiv& i \langle 0 \vert a^\dag_{\pb}(\tau)
  a_{\ka}(t)\vert 0 \rangle, \label{glkp} \\
  G^<_{ij}(t, \tau) &\equiv& i\langle 0 \vert a^\dag_j(\tau)
  a_i(t)\vert 0 \rangle. \label{glij}
\en where $\beta = L$ or $R$, and $\pb$ denotes a single-electron
state in the lead $\beta$. Hereafter we only solve the Green's
functions for time variables $t$ and $\tau$ ranging from $t_0^+$
to $+\infty$.
Taking the first-order time derivatives of $\grk$ and $\gak$ leads
to \be
  \left[i\partial_t - \epsilon_{\ka}(t) \right]\grk &=&
  \delta(t-\tau) \label{eom-grka}, \\
  -\left[i\partial_\tau + \eka(\tau)\right] g^a_{\ka}(t, \tau) &=&
  \delta(t - \tau), \label{eom-gaka}
\en with the initial conditions for Eq.~(\ref{eom-grka}):
$\grk\vert_{t = \tau^+} = -i$, $\grk\vert_{t = \tau} =
-\frac{i}{2}$, and $\grk\vert_{t = \tau^-} = 0$; and for
Eq.~(\ref{eom-gaka}): $g^a_{\ka}(t, \tau)\vert_{\tau = t^-} = 0$,
$g^a_{\ka}(t, \tau)\vert_{\tau = t} = \frac{i}{2}$, and
$g^a_{\ka}(t, \tau)\vert_{\tau = t^+} = i$, respectively. $\grk$
and $\gak$ can thus be utilized to solve partial differential and
integro-differential equations.
For instance, we have the EOM for $G^<_{\ka j}(t, \tau)$ as
follows, \beq
  \left[i\partial_t - \eka(t)\right] G^<_{\ka j}(t, \tau) = \sum_{m\in D}
  h_{\ka m}(t)\,G^<_{mj}(t, \tau). \label{eom-glkj-t-1}
\enq Combining Eqs.~(\ref{eom-glkj-t-1}) and (\ref{eom-grka}), we
obtain \be
  G^<_{\ka j}(t, \tau) &=& \sum_{m\in D}\int_{t_0^+}^t d\bar{t}\,
  g^r_{\ka}(t, \bar{t})\, h_{\ka m}(\bar{t})\, G^<_{mj}(\bar{t}, \tau)
  \nonumber \\
  && +\, i\, g^r_{\ka}(t, t_0)\, G^<_{\ka j}(t_0^+, \tau).
  \label{glkj-t}
\en With a similar but slightly more tedious treatment for the
time variable $\tau$, we arrive at \be
  G^<_{\ka j}(t, \tau) &=& -i\sum_{m\in D}G^<_{\ka m}(t,
  t_0^+)\, G^a_{mj}(t_0, \tau) \nonumber \\
  && -i\sum_{\pb}\sum_{m\in D}\int_{t_0^+}^\tau
  d\bar{t}\, G^<_{\ka \pb}(t, t_0^+) \nonumber \\
  && \times\, g^a_{\pb}(t_0, \bar{t})\, h_{\pb m}(\bar{t})
  \, G^a_{mj}(\bar{t}, \tau), \label{eqc}
\en where $\pb$ is short for $\sum_{\beta=L,R}\sum_{p\in\beta}$.
By taking $t = t_0^+$ in Eq.~(\ref{eqc}) and then insert it into
Eq.~(\ref{glkj-t}), we have \be
  G^<_{\ka j}(t, \tau) &=& \sum_{m\in D}\int_{t_0^+}^t d\bar{t}\, g^r_{\ka}(t,
  \bar{t})\, h_{\ka m}(\bar{t})\, G^<_{mj}(\bar{t}, \tau)
  \nonumber \\
  && +\, i\sum_{m\in D}g^r_{\ka}(t, t_0)\,\sigma_{\ka
  m}(t_0^+)\,  G^a_{mj}(t_0, \tau) \nonumber \\
  && +\, i\sum_{\pb}\sum_{m\in D}\int_{t_0^+}^\tau d\bar{t}\,
  g^r_{\ka}(t, t_0)\,\sigma_{\ka \pb}(t_0^+) \nonumber \\
  && \times \, g^a_{\pb}(t_0, \bar{t})\,h_{\pb
  m}(\bar{t})\,G^a_{mj}(\bar{t}, \tau), \label{eqd}
\en where the following equalities have been adopted: \be
  \sigma_{\ka m}(t) &=& -iG^<_{\ka m}(t, \tau)\Big\vert_{\tau = t}, \\
  \sigma_{\ka \pb}(t) &=& -iG^<_{\ka \pb}(t, \tau)\Big\vert_{\tau = t}.
\en
From Eq.~(\ref{qterm0}) the dissipative term $\iqa(t)$ is
expressed by \be
  Q_{\alpha, ij}(t) &=& i\sum_{k\in\alpha}h_{i\ka}(t)\, \sigma_{\ka j}(t) +
  \mbox{H.c.} \nonumber \\
  &=& \sum_{k\in\alpha}h_{i\ka}(t)\,G^<_{\ka j}(t, \tau)\vert_{\tau = t}
  + \mbox{H.c.}. \label{qa1}
\en Combining Eqs.~(\ref{eqd}) and (\ref{qa1}), we have thus \be
  Q_{\alpha, ij}(t) &=& \Bigg\{ Q^0_{\alpha, ij}(t) + \sum_{m\in
  D}\int_{t_0^+}^t d\bar{t}\, \Sigma^<_{\alpha, im}(t,
  \bar{t})\, G^a_{mj}(\bar{t}, t) \nonumber \\
  && +\, \sum_{m\in D}\int_{t_0^+}^t d\bar{t}\, \Sigma^r_{\alpha, im}(t,
  \bar{t})\, G^<_{mj}(\bar{t}, t) \Bigg\} \nonumber \\
  && +\, \mbox{H.c.}, \label{qa2}
\en where \be
  Q^0_{\alpha, ij}(t) &\equiv& i \sum_{k\in\alpha}\sum_{m\in D}h_{i\ka}(t)
  \, g^r_{\ka}(t, t_0)\, \sigma_{\ka m}(t_0^+) \nonumber \\
  && \times \, G^a_{mj}(t_0, t), \\
  \Sigma^<_{\alpha, im}(t, \bar{t}) &\equiv& i \sum_{k\in\alpha}\sum_{\pb}
  h_{i\ka}(t)\, g^r_{\ka}(t, t_0)\, \sigma_{\ka \pb}(t_0^+) \nonumber \\
  && \times \,g^a_{\pb}(t_0, \bar{t})\,h_{\pb m}(\bar{t}), \\
  \Sigma^r_{\alpha, im}(t, \bar{t}) &\equiv& \sum_{k\in\alpha} h_{i\ka}(t)\, g^r_{\ka}(t,
  \bar{t}) \, h_{\ka m}(\bar{t}),
\en where $\Sigma^<_\alpha$ and $\Sigma^r_\alpha$ are the lesser
and retarded self-energies of the device region, respectively.
Note that by definitions $G^r_D = [G^a_D]^\dag$, $G^<_D =
-[G^<_D]^\dag$, $\saa = [\Sigma^r_\alpha]^\dag$ and $\sla =
-[\sla]^\dag$, therefore, it's trivial to validate Eq.~(\ref{qa2})
is equivalent to Eq.~(\ref{e4qt2}).

It is important to emphasize that $\Psi(t_0^+)$ may be different
from $\Psi(t_0)$, so that the corresponding reduced
single-electron density matrix $\sigma(t_0^+)$ may also differ
from $\sigma(t_0)$. This would happen if
$\int_{t_0}^{t_0^+}H(\tau)\,d\tau \neq 0$, for instance, in the
cases where the external field involves a Delta function switched
on at $t_0$. However, for real physical systems, the applied
external field is real analytic in time. In this circumstance,
$\int_{t_0}^{t_0^+}H(\tau)\,d\tau = 0$, $\Psi(t_0^+) = \Psi(t_0)$,
and $\sigma(t_0^+) = \sigma(t_0)$.

The above derivations follow rigorously the partition-free scheme,
since the initial state $\Psi(t_0)$ can be the ground state of the
fully connected entire system including the device region and the
leads. As for the partitioned scheme, we need to introduce another
reference state $\Phi_0$, which is the ground state of Hamiltonian
$\tilde{H}$, \beq
  \tilde{H} \equiv \sum_{mn\in D}h_{mn}(t_0)a_m^\dag a_n +
  \sum_{\ka}\eka(t_0)a^\dag_{\ka}a_{\ka}. \label{hzero}
\enq Since $\tilde{H}$ does not contain any coupling terms between
$D$ and $L$ or $R$, $\Phi_0$ depicts the scenario that the device
region and the leads are isolated from each other. Hence there is
no electron populated across the boundary $S_L$ and $S_R$.
\emph{i.e.}, $\tilde\sigma_{D\alpha} = 0$ and $\tilde\sigma_{LR} =
0$. We now assume $\Psi(t_0)$ can be reached by a time propagation
of the entire system starting from the state $\Phi_0$,
\emph{i.e.}, \beq
  \Psi(t_0) = \mbox{e}^{-i\int_{-\infty}^{t_0}H(\tau)d\tau}\Phi_0.
  \label{psi-phi}
\enq At $t = -\infty$, $H(-\infty) = \tilde{H}$ and
$\sigma(-\infty) = \tilde\sigma$. In this sense, the initial time
for the Heisenberg creation and annihilation operators defined in
Eq.~(\ref{a-adag}) becomes $-\infty$ instead of $t_0$, and the
above derivations for the various Green's functions remain valid.
Note that for the decoupled ground state $\Phi_0$, we have \be
   \sigma_{ij}(-\infty) &=& \sigma^0_{ij}, \\
   \sigma_{\ka j}(-\infty) &=& 0, \\
   \sigma_{\ka \pb}(-\infty) &=&
   \delta_{\alpha\beta}\, \delta_{kp}\, f^0_{\ka},
\en where $f^0_{\ka}$ is the initial occupation number of the
single-electron state $\ka$. Thus the Green's functions and
self-energies previously derived can be simplified as follows, \be
  \Sigma^<_{\alpha, im}(t, \bar{t}) &=& i \sum_{k\in\alpha}
  f^0_{\ka}\,h_{i\ka}(t)\, g^r_{\ka}(t, t_0)\,
  g^a_{\pb}(t_0, \bar{t})\,h_{\pb m}(\bar{t}), \nonumber \\
  &=& \sum_{k\in\alpha} h_{i\ka}(t)\,g^<_{\ka}(t, \bar{t})\,h_{\ka
  m}(\bar{t}), \label{eqa1} \\
  G^<_{\ka j}(t, \tau) &=& \sum_{m\in D} \bigg\{ \int_{-\infty}^\tau
  d\bar{t}\,g^<_{\ka}(t, \bar{t})\,h_{\ka m}(\bar{t})\,G^a_{mj}(\bar{t},
  \tau) \nonumber \\
  &&+\int_{-\infty}^t \!\! d\bar{t}\, g^r_{\ka}(t, \bar{t})\,
  h_{\ka m}(\bar{t})\, G^<_{mj}(\bar{t}, \tau)\bigg\}, \label{eqb1} \\
  G^<_{ij}(t, \tau) &=& i\sum_{mn\in D}G^r_{im}(t,
  -\infty)\,\sigma^0_{mn}\,G^a_{nj}(-\infty, \tau) \nonumber \\
  && + \sum_{mn\in D}\int^t_{-\infty} dt_1 \int^\tau_{-\infty}
  dt_2\, G^r_{im}(t, t_1) \nonumber \\
  && \times \, \Sigma^<_{mn}(t_1, t_2)\, G^a_{nj}(t_2,
  \tau). \label{eqc1}
\en The dissipative term $\iqa$ is thus expressed as \be
  Q_{\alpha, ij}(t) &=& \sum_{m\in D}\int_{-\infty}^t
  d\bar{t}\,\Big[ \Sigma^<_{\alpha, im}(t, \bar{t})
  \,G^a_{mj}(\bar{t}, t) \nonumber \\
  &&  +\, \Sigma^r_{\alpha, im}(t, \bar{t})
  \,G^<_{mj}(\bar{t}, t) + \mbox{H.c.} \Big]. \label{qaij2}
\en Eqs.~(\ref{eqa1})$-$(\ref{qaij2}) recover exactly
Eq.~(\ref{e4qt}) derived from the Keldysh NEGF
formalism~\cite{keldysh}. Therefore, we conclude that as long as
the relation~(\ref{psi-phi}) holds, the partition-free and the
partitioned schemes of NEGF yield exactly the same dissipation
term $\iqa(t)$ for $t \geqslant t_0$.

In fact, Eq.~(\ref{psi-phi}) can be proved by Gell-Mann and Low
theorem (1951)~\cite{gml}, which basically states that $\Psi(t_0)$
can be reached from $\Phi_0$ by adiabatically turning on the
coupling terms between $D$ and $L$ or $R$ from $t = -\infty$ to
$t_0$. The resulting $\Psi(t_0)$ is an eigenstate of the
Hamiltonian $H(t_0)$ and in most cases is the ground state.

\section{Derivation of Eq.~(\ref{q-phys}) \label{app-iqa2}}

The greater Green's function for the reduced system, $G^>_{ij}(t,
\tau)$, is defined as \beq
  G^>_{ij}(t, \tau) \equiv -i \langle a_i(t) a^\dag_j(\tau)
  \rangle.
\enq The advanced Green's function of the reduced system can thus
be expressed as \beq
  G^a_D(t, \tau) = -\vartheta(\tau - t) \left[ G^>_D(t,
  \tau) - G^<_D(t, \tau) \right].
\enq Similarly the retarded and advanced self-energies can be
associated with the greater and lesser self-energies as follows,
\beq
  \Sigma^{r,a}_{\alpha}(t, \tau) = \pm\, \vartheta(\pm\, t \mp \tau)
  \left[ \sga(t, \tau) - \sla(t, \tau) \right],
\enq where the greater self-energy $\sga(t, \tau)$ is defined as
\be
   \Sigma^>_{\alpha, ij}(t, \tau) & \equiv & \sum_{k\in\alpha}
   h_{i\ka}(t)\,g^>_{\ka}(t, \tau)\, h_{\ka j}(\tau) \nonumber \\
   &=&-i\sum_{k\in\alpha}h_{i\ka}(t)\,h_{\ka j}(\tau)\, \nonumber
   \\
   && \times\, \langle \alpha \vert b_{\ka}(\tau) b^\dag_{\ka}(t)
   \vert \alpha \rangle.
\en Eq.~(\ref{e4qt}) is thus equivalent to \be
  \iqa(t) &=& \int^t_{-\infty} d\tau \left[ \Sigma^>_\alpha(t,
  \tau)\, G^<_D(\tau, t) \right. \nonumber \\
  &&  -\, \left.\Sigma^<_\alpha(t, \tau)\, G^>_D(\tau, t) \right] +
  \mbox{H.c.}. \label{qa-lg}
\en
In cases where the KS Fock matrix of the reduced system, $h_D$, is
time-independent, the greater and lesser Green's functions can be
approximated by QDT perturbatively to complete second
order~\cite{qdt}, \emph{i.e.}, \be \label{ggl-cs}
   G^>_D(\tau, t) & \approx & \mbox{e}^{i h_D (t - \tau)}
   G^>_D(t, t) = (-i)\,\mbox{e}^{i h_D (t - \tau)}\,\bar{\sigma}_D,
   \nonumber \\
   G^<_D(\tau, t) & \approx & \mbox{e}^{i h_D (t - \tau)}
   G^<_D(t, t) = i\,\mbox{e}^{i h_D (t - \tau)}\, \sigma_D,
\en where $\bar{\sigma}_D \equiv I - \sigma_D$ is the reduced
single-hole density matrix of the reduced system. Assuming the
lead Hamiltonian to be time-independent, we have
$\Sigma_\alpha^{<,>}(t, \tau) = \Sigma_\alpha^{<,>}(t - \tau)$.
Hence, Eq.~(\ref{qa-lg}) can be recast into \be
   \iqa(t) &=& \int_0^\infty d\tau\, \sla(\tau)\, \mbox{e}^{i h_D \tau}
   G^>_D(t, t)  \nonumber \\
   && + \int_0^\infty d\tau\, \sga(\tau)\, \mbox{e}^{i h_D
   \tau} G^<_D(t, t) + \mbox{H.c.}. \label{qterm-a}
\en To evaluate the causality transforms involved in
Eq.~(\ref{qterm-a}), we define \be
   \Lambda^{(\pm)}_\alpha(h_D) & \equiv & \pm \frac{1}{2i} \int_0^\infty d\tau
   \left[ \Sigma^{<,>}_\alpha(\tau)\,\mbox{e}^{i h_D \tau} \right.\nonumber \\
   && + \left. \mbox{e}^{-i h_D \tau}
   \Sigma^{<,>}_\alpha(-\tau) \right], \label{lambda-pm} \\
   \Gamma^{(\pm)}_\alpha(h_D) & \equiv & \mp \frac{1}{2} \int_0^\infty d\tau
   \left[ \Sigma^{<,>}_\alpha(\tau)\,\mbox{e}^{i h_D \tau} \right.\nonumber \\
   && - \left. \mbox{e}^{-i h_D \tau}
   \Sigma^{<,>}_\alpha(-\tau) \right]. \label{gamma-pm}
\en Here the equality $[\Sigma^{<,>}_\alpha(\tau)]^\dag =
-\Sigma^{<,>}_\alpha(-\tau)$ has been adopted. With
Eqs.~(\ref{qterm-a})$-$(\ref{gamma-pm}),
Eqs.~(\ref{q-phys})$-$(\ref{q-phys2}) are readily recovered.
Generally $\Lambda^{(\pm)}_\alpha(h_D)$ and
$\Gamma^{(\pm)}_\alpha(h_D)$ are Hermitian matrices, and
associated with each other via the Kramers-Kronig
relation~\cite{qdt}. In particular, when the KS Fock matrix $h$ is
real, $\Lambda^{(\pm)}_\alpha(h_D)$ and
$\Gamma^{(\pm)}_\alpha(h_D)$ become real symmetric matrices.
With $\iqa$ expressed by Eq.~(\ref{q-phys}), the EOM for
$\sigma_D$ is reformulated as
\be \label{eom-phys}
   i\dot{\sigma}_D &=& [h_D, \sigma_D] + \sum_{\alpha=L,R}
   [\tilde{\Sigma}^>_{\alpha}(h_D), \sigma_D]^\dag \nonumber \\
   && + \sum_{\alpha=L,R}
   [\tilde{\Sigma}^<_{\alpha}(h_D), \bar{\sigma}_D]^\dag.
\en
Eq.~(\ref{eom-phys}) resembles closely Eq.~(8) in
Ref.~\cite{csqdt-scba}, which is developed from CS-QDT with the
Markovian approximation. The correlation functions of the leads
used in Ref.~\cite{csqdt-scba}, $C^{(\pm)}_\alpha(t, \tau)$, are
related to the self-energies adopted in our work as follows, \beq
   \Sigma^{<,>}_\alpha(t, \tau) = \pm\, i\, C^{(\pm)}_\alpha(t, \tau).
\enq
Following the SCBA scheme proposed in Ref.~\cite{csqdt-scba},
higher order effects due to interactions between the reduced
system and the environment can be partially accounted for by
substituting in Eq.~(\ref{ggl-cs}) an effective propagator of the
reduced system, $\mbox{e}^{i h^{\mathit{eff}}_D(t - \tau)}$, for
the propagator of the isolated reduced system, $\mbox{e}^{i h_D(t
- \tau)}$, where $h^{\mathit{eff}}_D$ is some effective KS Fock
matrix of the reduced system. This results in self-energy terms
$\tilde{\Sigma}^{<,>}_{\alpha}(h^{\mathit{eff}}_D)$ instead of
$\tilde{\Sigma}^{<,>}_{\alpha}(h_D)$ in Eq.~(\ref{eom-phys}).
\section{Wide-band limit scheme for the dissipation
term $\mathbf{Q}_{\mathbf{\alpha}}$ \label{EOM4WBL}}

With the WBL approximation, the advanced self-energy becomes local
in time~\cite{prb94win},
\begin{eqnarray}
    \Sigma^{a}_{\alpha,nm}(\tau,t) &=& \sum_{k_{\alpha}\in\alpha}
    h_{n\ka}(\tau)\, h_{\ka m}(t)\, g^a_{\ka}(\tau, t)
    \nonumber \\
    &=& \sum_{\ka} h_{n\ka}(\tau)\, h_{\ka m}(t)\nonumber \\
    &&\times\,\left[\,i\vartheta(t-\tau)\,
    \mbox{e}^{\,i\epsilon _{k}^{\alpha}(t-\tau)}\,\mbox{e}^
    {\,i\int^{t}_{\tau}\Delta\epsilon^{\alpha}(\bar{t})\,d\bar{t}}\,\right]
    \nonumber \\
    &=& \frac{i}{\pi}\,\vartheta
    (t-\tau)\,\mbox{e}^{\,i\int^{t}_{\tau}\Delta\epsilon^{\alpha}
    (\bar{t})\,d\bar{t}} \nonumber\\
    &&\times\left\{
    \int_{-\infty}^{+\infty}\mbox{e}^{i\epsilon(t-\tau)}
    d\epsilon\right\}\Lambda^{\alpha}_{nm} \nonumber \\
    &=& i\delta(t-\tau)\Lambda^{\alpha}_{nm}. \label{aselfe}
\end{eqnarray}
Here the Dirac Delta function on the RHS effectively removes the
tricky off-diagonal elements of $G^<_D(t, \tau)$ from the NEGF
formulation for $\iqa$ (\emph{cf.} Eq.~(\ref{e4qt})). The third
equality of Eq.~(\ref{aselfe}) involves the following
approximation for the line-widths within the WBL approximation,
\begin{eqnarray}
   \Lambda^{\alpha}_{k, nm}(t,\tau) & \equiv & \pi\,\eta_{\alpha}
   (\epsilon^{\alpha}_k)\,h_{nk_ {\alpha}}(t)\,h_{k_{\alpha}m}(\tau)
   \nonumber \\
   & \approx & \Lambda_{nm}^{\alpha}(t,\tau) \approx
   \Lambda_{nm}^\alpha. \label{lambda}
\end{eqnarray}
At time $t = 0$ the entire fully connected system ($D$ + $L$ +
$R$) is in its ground state with the chemical potential $\mu^0$.
Afterwards the external potential is switched on, resulting in
homogeneous time-dependent level shifts $\dea(t)$ for the lead
$\alpha$ ($L$ or $R$). Hence, for $t,\tau>0$ we have
\begin{eqnarray}\label{tau-gt-zero}
    \Sigma^{<}_{\alpha,nm}(\tau,t) &=& \sum_{k_{\alpha}\in\alpha}
    h_{n\ka}(\tau)\, h_{\ka m}(t)\, g^{<}_{\ka}(\tau,t) \nonumber \\
    &=&\sum_{\ka\in\alpha}h_{n\ka}(\tau)\, h_{\ka m}(t) \nonumber\\
    &&\times\left[i\,f^{\alpha}(\epsilon^{\alpha}_{k})\,
    \mbox{e}^{\,i\epsilon _{k}^{\alpha}(t-\tau)}\,\mbox{e}^
    {\,i\int^{t}_{\tau}\Delta\epsilon^{\alpha}(\bar{t})\,d\bar{t}}
    \,\right]\nonumber\\
    &=& \frac{2i}{\pi}\,\mbox{e}^{\,i\int^{t}_{\tau}\Delta\epsilon^{\alpha}
    (\bar{t})\,d\bar{t}}\,\Lambda^{\alpha}_{nm}\nonumber\\
    &&\times\left\{\int_{-\infty}^{+\infty}f^{\alpha}(\epsilon)\,\mbox{e}^
    {\,i\epsilon(t-\tau)}d\epsilon\right\}, \label{lselfe-gt-zero}\\
    G^{r}_{nm}(t,\tau)&=&-i\vartheta(t-\tau)\sum_{l\in D}U^{(-)}_{nl}
    (t)U^{(+)}_{lm}(\tau),\label{gr4wbl-gt-zero}
\end{eqnarray}
while for $\tau<0$ and $t > 0$, the counterparts of
(\ref{lselfe-gt-zero}) and (\ref{gr4wbl-gt-zero}) are as follows,
\begin{eqnarray}
    \Sigma^{<}_{\alpha,nm}(\tau,t)&=&\sum_{k_{\alpha}\in\alpha}
    h_{nk_{\alpha}}(\tau)\, h_{k_{\alpha}m}(t)\, g^{<}_{k_{\alpha}}(\tau,t)
    \nonumber \\
    &=&\sum_{k_{\alpha}\in\alpha}h_{nk_{\alpha}}(\tau)\, h_{k_{\alpha}m}(t)
    \nonumber \\
    &&\times\,\Big[i\,f^{\alpha}(\epsilon^{\alpha}_{k})\,\mbox{e}^{\,i
    \epsilon^{\alpha}_{k}(t-\tau)}\mbox{e}^{\,i\int^{t}_{0}\Delta
    \epsilon^{\alpha}(\bar{t})\,d\bar{t}}\,\Big] \nonumber\\
    &=& \frac{2i}{\pi}\,\mbox{e}^{\,i\int^{t}_{0}\Delta
    \epsilon^{\alpha}(\bar{t})\,d\bar{t}}
    \,\Lambda^{\alpha}_{nm} \nonumber \\
    && \times\,\left\{\int_{-\infty}^{+\infty}f^{\alpha}(\epsilon)\,\mbox{e}^
    {\,i\epsilon(t-\tau)}d\epsilon\right\}, \label{lselfe-lt-zero}
\end{eqnarray}
\beq
    G^{r}_{nm}(t,\tau) = \sum_{l\in D}U^{(-)}_{nl}(t)\,G^{r}_{lm}
    (0,\tau).  \label{gr4wbl-lt-zero}
\enq Here the effective propagators for the reduced system,
$U^{(\pm)}(t)$, are defined as
\begin{eqnarray}\label{u+u-0}
    U^{(\pm)}(t)&=&\exp\bigg\{\pm i\int_{0}^{t}\!h_{D}(\tau)d\tau
    \pm\Lambda\,t\bigg\},
\end{eqnarray}
where $\Lambda = \sum_{\alpha=L,R}\Lambda^{\alpha}$. By inserting
Eqs.~(\ref{aselfe})$-$(\ref{gr4wbl-lt-zero}) into
Eq.~(\ref{e4qt}), the dissipation term $Q_\alpha$ is simplified to
be
\begin{eqnarray}
    \iqawbl(t) &=& K^{\alpha}(t) +
    \left\{\Lambda^{\alpha},\sigma_{D}(t)\right\}, \label{qfinal0}
\end{eqnarray}
where the curly bracket on the RHS denotes an anticommutator, and
$K^{\alpha}(t)$ is a Hermitian matrix:
\begin{eqnarray}
    K^{\alpha}(t) &=& P^{\alpha}(t) + \left[P^{\alpha}(t)\right]
    ^{\dagger}. \label{k-alpha-wbl}
\end{eqnarray}
Here $P^{\alpha}(t)$ involves an integration over the entire real
$t$-axis, which is then decomposed into positive and negative
parts, denoted by $P^{(+)}_{\alpha}(t)$ and $P^{(-)}_{\alpha}(t)$,
respectively. We thus have
\begin{eqnarray}
    P^{\alpha}(t) & \equiv & -\int_{-\infty}^{+\infty}d\tau\,
    G^{r}_{D}(t,\tau)\Sigma^{<}_{\alpha}(\tau,t) \nonumber \\
    &=& P_{\alpha}^{(-)}(t) + P_{\alpha}^{(+)}(t).\label{p-alpha}
\end{eqnarray}
$P_{\alpha}^{(-)}(t)$ and $P_{\alpha}^{(+)}(t)$ are evaluated via
\begin{eqnarray}\label{p-alpha-minus}
    P_{\alpha}^{(-)}(t) & \equiv & -\int_{-\infty}^{0}d\tau\,
    G^{r}_{D}(t,\tau)\Sigma^{<}_{\alpha}(\tau,t)\nonumber\\
    &=&-\frac{2i}{\pi}\,\,\mbox{exp}\left\{i\!\int_{0}^{t}\Delta
    \epsilon^{\alpha}\!(\tau)d\tau\right\}U^{(-)}(t)
    \nonumber \\
    && \times \left\{\int_{-\infty}^{\mu^{0}}\frac
    {d\epsilon\,\mbox{e}^{i\epsilon t}}{\epsilon-h_{D}(0)+i\,\Lambda}
    \right\}\Lambda^{\alpha},
\end{eqnarray}
and
\begin{eqnarray} \label{p_alpha_plus_2}
   P_{\alpha}^{(+)}(t) & \equiv & -\frac{2}{\pi}\int^{\mu^0}_{-\infty}
   d\epsilon\,W_{\alpha}^{(-)}(\epsilon,t) \nonumber \\
   && \times \int^{t}_{0}d\tau\,
   W_{\alpha}^{(+)}(\epsilon,\tau)\,\Lambda^{\alpha},
\end{eqnarray}
respectively, where
\begin{eqnarray} \label{w_alpha_plus}
   W_{\alpha}^{\pm}(\epsilon,t) &=& \mbox{e}^{ \pm\,i
   \int^{t}_{0}d\tau \left[h_{D}(\tau) -i\Lambda -
   \Delta\epsilon^{\alpha}(\tau) - \epsilon\right]}.
\end{eqnarray}
However, the evaluations of
Eqs.~(\ref{p_alpha_plus_2})$-$(\ref{w_alpha_plus}) are found
extremely time-consuming since at every time $t$ one needs to
propagate $W_{\alpha}^{\pm}(\epsilon,t)$ for every individual
$\epsilon$ inside the lead energy spectrum. It is thus inevitable
to have a simpler approximate form for $P^{(+)}_{\alpha}(t)$ with
satisfactory accuracy retained. Note that
Eq.~(\ref{p_alpha_plus_2}) can be reformulated as
\begin{eqnarray}
   \ppt &=& -\frac{2}{\pi}\int_{-\infty}^{\mu^0}d\epsilon
   \int_0^t d\tau \nonumber \\
   && \times\,\mbox{e}^{-i\int_{\tau}^t \left[h_D(\bar{t})-i\Lambda
   -\dea(\bar{t})-\epsilon \right]d\bar{t}}\,\Lambda^{\alpha}.
   \label{paplus1}
\end{eqnarray}
For cases where steady states can be ultimately reached,
$\Delta\epsilon^{\alpha}(t)$ and $h_D(t)$ become asymptotically
constant as time $t\rightarrow +\infty$, \emph{i.e.},
$\dea(t)\rightarrow \dea(\infty)$ and $h_D(t)\rightarrow
h_D(\infty)$. Therefore, the steady state
$P_{\alpha}^{(+)}(\infty)$ can be approximated by substituting
$\dea(\infty)$ and $h_D(\infty)$ for $\dea(t)$ and $h_D(t)$ in
Eq.~(\ref{paplus1}), respectively.
\begin{eqnarray}
   P_{\alpha}^{(+)}(\infty) & \approx &
   -\frac{2}{\pi}\int_{-\infty}^{\mu^0}d\epsilon \int_0^t d\tau\ \nonumber \\
   && \times\,\mbox{e}^{-i\left[h_D(\infty) -i\Lambda
   -\Delta\epsilon^{\alpha}(\infty)-\epsilon \right](t-\tau)}
   \,\Lambda^{\alpha}\nonumber \\
   &=& -\frac{2i}{\pi}\int_{-\infty}^{\mu^0} \left\{I -
   \mbox{e}^{-i\left[h_D(\infty) -i\Lambda -\dea(\infty) - \epsilon
   \right]t} \right\} \nonumber\\
   && \times\,\frac{d\epsilon}{\epsilon - h_D(\infty) + i\Lambda + \dea(\infty)}
   \,\Lambda^{\alpha}.  \label{paplus2}
\end{eqnarray}
It is obvious from Eq.~(\ref{paplus1}) that
\begin{eqnarray}
 P_{\alpha}^{(+)}(0) = 0. \label{paplus10}
\end{eqnarray}
Thus $\ppt$ for any time $t$ between $0$ and $+\infty$ can be
approximately  expressed by adiabatically connecting Eq.~(\ref{paplus2})
with (\ref{paplus10}) as follows,
\begin{eqnarray}
   P_{\alpha}^{(+)}(t) & \approx &
   -\frac{2i}{\pi}\int_{-\infty}^{\mu^0} \left\{I -
   \mbox{e}^{-i\int_0^t \left[h_D(\tau) -i\Lambda
   -\dea(\tau) - \epsilon\right]d\tau} \right\} \nonumber\\
   && \times\,\frac{d\epsilon}{\epsilon -
   h_D(t) + i\Lambda + \dea(t)}\,\Lambda^{\alpha}.
   \label{paplus3}
\end{eqnarray}
Both Eqs.~(\ref{paplus1}) and (\ref{paplus3}) lead to the correct
$P^{\alpha}(\infty)$ for steady states,
\begin{eqnarray}
   P^{\alpha}(\infty) &=& -\frac{2i}{\pi}\!\int_{-\infty}^{\mu^0}
   \!d\epsilon \nonumber \\
   && \times\,\frac{1}{\epsilon - h_{D}(\infty) + i \Lambda +
   \Delta\epsilon^{\alpha}(\infty)}\,\Lambda^{\alpha}, \label{pwblss}
\end{eqnarray}
If the external applied voltage assumes a step-like form, for
instance, $\Delta V^{\alpha}(t) = -\dea(t) = \Delta V^{\alpha}(1 -
\mbox{e}^{-t / a})$ with $a\rightarrow 0^+$, and $h_D(t)$ is not
affected by the fluctuation of $\sigma_D(t)$, Eq.~(\ref{paplus3})
would recover exactly Eq.~(\ref{paplus1}). In other cases,
Eq.~(\ref{paplus3}) provides an accurate and efficient
approximation for Eq.~(\ref{paplus1}), so long as $\Delta
V^{\alpha}(t)$ do not vary dramatically in time. Since the
integration over energy in Eq.~(\ref{paplus3}) can be performed
readily by transforming the integrand into a diagonal
representation, Eq.~(\ref{paplus3}) is evaluated much faster than
Eq.~(\ref{paplus1}). Due to its efficiency and accuracy,
Eq.~(\ref{paplus3}) is then combined with
Eqs.~(\ref{qfinal0})$-$(\ref{p-alpha-minus}) to calculate the
dissipation term $\iqawbl$, and thus recovers Eq.~(\ref{k-wbl-1})
of Sec.~\ref{schemes}.

\end{document}